\documentclass{aa}  
\usepackage{graphicx}
\usepackage{txfonts,textcomp}
\usepackage{hyperref}

\newcommand{\los}[1]{$los$}
\newcommand{\cii}[1]{$\rm [C_{II}]$}
\newcommand{\ci}[1]{$\rm [C_{I}]$}
\newcommand{\ciidl}[1]{$\rm [C_{II}]_{(DL14)}$}
\newcommand{\ciila}[1]{$\rm [C_{II}]_{(L18)}$}
\newcommand{\poker}{POKER}

\newcommand{\zcii}[1]{$z_{\rm [C_{II}]}$}
\newcommand{\karcmin}[1]{$\rm arcmin^{-1}$}
\newcommand{\rsigma}[1]{$r_{\rm \sigma}$}

\begin{document}

   \title{CONCERTO: Extracting the power spectrum of the \cii \, emission line}


   \author{M. Van Cuyck \inst{1} \and
          N. Ponthieu \inst{2,1} \and  
          G. Lagache \inst{1} \and
          A. Beelen \inst{1} \and
          M. Béthermin \inst{1, 3} \and
          A. Gkogkou \inst{1} \and
          M. Aravena \inst{4} \and
          A. Benoit \inst{5} \and
          J. Bounmy \inst{6} \and
          M. Calvo \inst{5} \and
          A. Catalano \inst{6} \and
          F.X. Désert \inst{2} \and
          F.-X. Dupé \inst{1} \and
          A. Fasano \inst{1} \and 
          A. Ferrara \inst{7} \and
          J. Goupy  \inst{5} \and
          C. Hoarau  \inst{6} \and
          W. Hu \inst{1} \and 
          J.-C Lambert \inst{1} \and
          J. F. Mac\'ias-P\'erez \inst{6} \and
          J. Marpaud \inst{6} \and
          G. Mellema \inst{8} \and
          A. Monfardini \inst{5} \and
          A. Pallottini \inst{7}
          }

   \institute{Aix Marseille Univ, CNRS, CNES, LAM, Marseille, France, 
             \and
             Univ. Grenoble Alpes, CNRS, IPAG, 38000 Grenoble, France, 
             \and
             Université de Strasbourg, CNRS, Observatoire astronomique de Strasbourg, UMR 7550, 67000 Strasbourg, France  
             \and
             Núcleo de Astronomía, Facultad de Ingeniería y Ciencias, Universidad Diego Portales, Av. Ejército 441, Santiago, Chile,
             \and
             Univ. Grenoble Alpes, CNRS, Grenoble INP, Institut Néel, 38000 Grenoble, France
             \and 
             Univ. Grenoble Alpes, CNRS, LPSC/IN2P3, 38000 Grenoble, France,
             \and
             Scuola Normale Superiore, Piazza dei Cavalieri 7, 56126 Pisa, Italy
             \and 
             The Oskar Klein Centre, Department of Astronomy, Stockholm University, AlbaNova, SE-10691 Stockholm, Sweden
             }

   \date{Received Month day, year; accepted month day, year}

  \abstract
   {CONCERTO is the first experiment to perform a \cii\, line intensity mapping (LIM) survey on the COSMOS field to target $z>5.2$. Measuring the \cii \, angular power spectrum allows us to study the role of dusty star-forming galaxies in the star formation history during the epochs of Reionization and post-Reionization. The main obstacle to this measurement is the contamination by bright foregrounds: the dust continuum emission and atomic and molecular lines from foreground galaxies at $z \lesssim 3$.}
   {We evaluate our ability to retrieve the \cii\, signal in mock observations of the sky using the Simulated Infrared Dusty Extragalactic Sky (SIDES), which covers covering the mid-infrared to millimetre range. We also measure the impact of field-to-field variance on the residual foreground contamination.}
   {We compared two methods for dealing with the dust continuum emission from galaxies (i.e. the cosmic infrared background fluctuations): the standard principal component analysis (PCA) and the asymmetric re-weighted penalized least-squares (arPLS) method. For line interlopers, the strategy relies on masking low-redshift galaxies using the instrumental beam profile and external catalogues. As we do not have observations of CO or deep-enough classical CO proxies (such as $L_{\rm IR}$), we relied on the COSMOS stellar mass catalogue, which we demonstrate to be a reliable CO proxy for masking. To measure the angular power spectrum of masked data, we adapted the P of K EstimatoR (\poker) from cosmic infrared background studies and discuss its use on LIM data.}
   {The arPLS method achieves a reduction in the cosmic infrared background fluctuations to a sub-dominant level of the \cii\, power at z$\sim$7, a factor of $>70$ below our fiducial \cii\, model. When using the standard PCA, this factor is only 0.7 at this redshift. The masking lowers the power amplitude of line contamination down to $\rm 2 \times 10^{-2}$\,Jy$^2$/sr. This residual level is dominated by faint undetected sources that are not clustered around the detected (and masked) sources. For our \cii\, model, this results in a detection at $z=5.2$ with a power ratio \cii\,\,/\,(residual interlopers)\,=\,$62\pm 32$ for a 22\,$\%$ area survey loss. However, at $z=7$, \cii\,\,/\,(residual interlopers)\,=\,$2.0 \pm 1.4$, due to the weak contrast between \cii\, and the residual line contamination. Thanks to the large area covered by SIDES-Uchuu, we show that the power amplitude of line residuals varies by 12-15\,$\rm \%$ for z=5.2-7, which is less than the field-to-field variance affecting \cii\, power spectra.}
   {We present an end-to-end simulation of the extragalactic foreground removal that we ran to detect the \cii\, at high redshift via its angular power spectrum. We show that cosmic infrared background fluctuations are not a limiting foreground for \cii\, LIM. On the contrary, the CO and \ci\, line contamination severely limits our ability to accurately measure the \cii\, angular power spectrum at $z\gtrsim 7$.}

   \maketitle

%

\section{ \label{introduction} Introduction} 
The Epoch of Reionization (EoR) started with the birth of the first stars and the formation of the first galaxies, whose ultraviolet light ionized the neutral hydrogen of the intergalactic medium. As time went by, the bubbles of ionized gas increased in number, expanding and merging, eventually reionizing the whole Universe. While we are beginning to understand the overall chronology of the EoR, the full story of reionization remains to be written. However, recent and planned observational advances promise progress in this respect over the coming decade.

The \textit{Planck} satellite has provided an accurate measurement of the Thomson scattering optical depth, $\tau$, using the cosmic microwave background (CMB) angular power spectrum \citep[APS;][]{planck_2016}. It shows that the Universe was reionized at 10$\%$ at $z_{\rm 10\%}=10.4 \pm 1.8$. On the other end of the reionization epoch, studies of the ionized intergalactic medium through quasar and gamma-ray burst absorption in the Lyman-$\rm \alpha$ and Lyman-$\rm \beta$ hydrogen transitions set the late end at $z \approx 6$ \citep[e.g.][]{chornock_grb, 2019ApJ...881...23E, 2020_wang_gp, bosman_eor}. As of today, the total ionizing photon budget required to reionize the Universe is still not probed by the deep surveys that have constrained the faint-end slope of the ultraviolet luminosity function in the early Universe \cite[e.g.][]{2013ApJ...763L...7E, hudf2, alma_hudf, hudf, 2017A&A...608A...6D}. This may change in the near future with deep observations from the\textit{ James Webb} Space Telescope (JWST). 
 
The CarbON CII line in post-rEionization and ReionizaTiOn (CONCERTO) project \citep{concerto_2020} seeks to constrain the dust-obscured star formation history and its spatial distribution at high redshifts (z>5.2). To that aim, CONCERTO is the first experiment to conduct a line intensity mapping (LIM) survey of the \cii\, line at the end of the EoR and during the post-reionization era. A LIM survey consists of mapping the surface brightness in a given field of view as a function of position and observed frequency (see \citealt{lim_report_2017} and \citealt{lim_report_2022} for a review). Targeting an emission line then allows the redshift information to be recovered. Line intensity mapping is naturally sensitive to faint sources, missed by classical galaxy surveys, that potentially contribute the most to the ionizing photon budget \cite[e.g.][]{2012ApJ...758...93F, 2013MNRAS.434.1486D, 2015ApJ...814...69A}. The CONCERTO project then studies surface brightness fluctuations in the Fourier space using the APS. 

The \cii\, 157.7 $\rm \mu$m line is a line of choice for LIM at high redshifts. It is one of the brightest far-infrared lines and is unattenuated by the dust content of galaxies. 
Furthermore, it has been shown by both observations \cite[e.g.][]{2015MNRAS.449.2883G, alpines} and theoretical works \cite[e.g.][]{2017MNRAS.471.4128P, 2018A&A...609A.130L} that \cii\, luminosity is a reliable tracer of star formation in the redshift range of interest. \cii\, emission from high-redshift galaxies is dominated by the emission from photo-dissociation regions, which are regions of cool neutral gas penetrated by ionizing ultraviolet radiations from hot young stars \citep{2014_pineda_cii} that surround molecular clouds \citep{vallini_2017, pallottini_2022}. In addition, the transparent millimetre atmospheric window enables access from the ground to the \cii\, emission originating at z$\gtrsim$4. 
The main challenge for any kind of LIM experiment remains the foreground contamination, which can completely dominate the signal of interest. For the CONCERTO \cii\, survey, the brightest foreground comes from fluctuations in the cosmic infrared background (CIB). The CIB is the line-of-sight-integrated emission of all dusty galaxies over the entire history of the Universe, from 8 to 1000\,$\rm \mu$m \citep{2000LNP...548...81L}. Further, there is a second foreground due to interloping emission lines, which are atomic and molecular emission lines redshifted in the same frequency band. For the \cii\, survey in the CONCERTO frequency range, the main interlopers are the CO rotational ladder and the two \ci\, fine structure lines from foreground galaxies at $z\lesssim3$. In this paper we present in detail a deconfusion method designed for the \cii\, survey conducted with CONCERTO. \\

The treatment of continuum foregrounds has mainly been studied in the context of 21\,cm LIM. Most of the techniques rely on both its spectral smoothness and its strong fluctuations to reconstruct and subtract it \citep{mcquinn_2006_21cm, furlanetto2006_continuum_21cm, morales_2006_continuum_21cm, chang_2010_crossco, parsons_2012_continuum_21cm, liu_et_tegmark_2012_continuum_21cm, switzer_2015_continuum, yue_2015_masking} or used to avoid the contaminated angular scales in the (3D) Fourier space \citep{chapman_2016_gmca}. 
Regarding the interlopers, the cross correlation between the data and an alternative tracer probing the same cosmic volume can be used to disentangle the signal of interest from the interlopers. This option has been used for popular lines in intensity mapping \citep{lidz_2009_crossco, visbal_2010_crossco, masui_2013_crossco, croft_2016_crossco, chung_2019_crossco} as well as for the \cii\, line \citep{gong_2012_crossco, switzer_2019_crossco, cross_cii_oiii, 2022_cii_co, 2022_pullen} below $z<3$. However, there are not enough ancillary probes tracing the EoR at $z>6$ to cross correlate with the \cii\, LIM data \citep[but see][]{chang_2015_crossco, 2016_comaschi_crossco_cii_lya}. A variety of deconfusion methods are thus under study. By exploiting the anisotropies in the APS of interlopers projected on the \cii\, frame \citep{cheng_2016_deconfusion, lidz_anisotropies, gong_2020_multipole}, it is possible to separate out the interloper contribution in the power spectrum, but a high sensitivity and a large-area survey are required to detect these anisotropy features. 
\citet{cheng_2020_phasespace} propose a deconfusion method in the phase space for reconstructing the one-point statistics of CO-contaminated data, which can only be used if the CO spectral line energy distribution (SLED) in galaxies varies by less than 20$\%$ and with a reasonable noise level. A deep learning approach for de-blending $\rm H_{\alpha}$ at $z=1.3$ from [$\rm O_{III}$] at $z=2$ was also investigated in \cite{moriwaki_component_deeplearning}; it achieved a 91\% precision but was not applied to the \cii\, case. \\

Among the interloper separation methods, masking is conceptually the simplest one. The \cii \, signal is not correlated with the signal from interlopers as they originate from lower redshifts. Hence, removing interloper-contaminated pixels should give an unbiased measurement of the \cii\, power spectrum by lowering the interlopers' amplitude in the total APS. 
Masking efficiency has been studied through its effect on the clustering and shot noise power spectrum levels \citep{gong_2014_masking, breysse_kovetz_masking_ciidifficult, sun_masking}. 
However, due to the lack of realistic far-infrared to millimetre sky simulations to apply this method to, many side effects are poorly studied, such as the correlation between faint and high-mass galaxies, the correlation between the signal and the mask, and the mixing of $k$ modes induced by the mask. \cite{yue_2015_masking} applied the masking to mock maps, but they did not investigate the systematics as cosmic variance or residual contamination.\\

In this paper we study all the steps for an end-to-end component separation method applied to the \cii\, LIM survey conducted with CONCERTO in order to extract the \cii\, APS. We are interested in the detection significance and accuracy of the \cii\, APS reconstruction that can be achieved in a typical CONCERTO field after foreground removal. We also study the effect of the field-to-field (FtF) variance on the masking results for CONCERTO-like field sizes. This variance affects both the interloper and the \cii\, signals.

For this, we used Simulated Infrared Dusty Extragalactic Sky (SIDES), a newly developed realistic empirical simulation of the dusty infrared-millimetre sky up to $z=7$, which is fully described in \citet{bet17, bet22} and \citet{gko22}, hereafter B17, B22, and G22, respectively. The component separation methods are applied directly to these mock observations, as they would be applied to the real CONCERTO observations. For the CIB contamination (continuum emission), our component separation is based on two different approaches. The first is principal component analysis (PCA) adapted to LIM data \citep[e.g.][]{bg_pca_2015, yohana_2021_pca}, and the second is a baseline estimator called asymmetric reweighed penalized least squares \citep[arPLS;][]{ARPLS}. For the interloper emission lines, as CONCERTO carries out the \cii\, survey on the widely studied Cosmic Evolution Survey \citep[COSMOS;][]{2007ApJS..172....1S} field, we rely on a custom-made masking strategy. In this observational context, our main focus is on the 2D APS, which is the initial quantity derived from the data. The APS of each species can be directly compared, unlike the spherically averaged power spectra, which require projection onto the \cii\, frame. Additionally, as detailed in B22, the lowest radial modes will not be explored, due to CONCERTO's low spectral resolution, and the majority of the captured signal will be predominantly present in the well-sampled transverse modes.

The paper is organized as follows. We first present the CONCERTO mock observation cubes from SIDES (Sect.\,\ref{simu}). In Sect.\,\ref{continuum removal} we address the removal of CIB anisotropy contamination. In Sect.\,\ref{sect:masking} we present the masking strategy based on the use of the stellar mass, $M_*$, as a CO proxy and a beam width criterion. In Sect.\,\ref{sect:masking_result} we present the masking results on the interloper maps alone and the interloper-contaminated \cii\, maps. Finally, we summarize our results and conclude in Sect.\,\ref{ccl}. The validation of our APS estimator is detailed in Appendix\,\ref{poker_appendix}. 
Throughout the paper we assume a \cite{refId0} cosmology and a \cite{2003PASP..115..763C} initial mass function, as in SIDES (B17).

\section{\label{simu} SIDES simulation} 

The SIDES simulation, which mimics the CONCERTO extragalactic sky, is a central element of our study. SIDES accurately replicates a series of statistical properties, spanning from mid-infrared to millimetre wavelengths, including the redshift distributions and number counts of galaxies, line luminosity functions, and CIB fluctuations (see B17, B22, and G22 for a complete description of the simulation and for comparisons with observational constraints). It produces realistic LIM mock data constituting our input sky used to evaluate the component separation method efficiency.

In Sect.\,\ref{sides_cat}, we briefly summarize the important SIDES characteristics for our use. We further review the map and cube-making process in Sect.\,\ref{making}. 

\subsection{The SIDES catalogue \label{sides_cat}} 
SIDES produces a catalogue of clustered galaxies with their sky coordinates and redshift (up to $z=7$), together with a large set of physical parameters, as for example their star formation rate (SFR), stellar mass ($M_*$), or flux in several instrument bandpasses. All relations used to get these physical parameters are derived empirically from observational trends and include an appropriate scatter. First, from a dark matter (DM) simulation, SIDES produces a galaxy light cone by linking the stellar mass $M_*$ to the DM halos' mass, using abundance matching. Two versions exist: one using the $1.4\times 1.4$\,deg$^2$ DM light cone from the Bolshoi-\textit{Planck} simulation (B17, B22) and one using the 117\,deg$^2$ DM light cone from the Uchuu simulation (G22). Next, galaxies are randomly assigned to star-forming or passive galaxies, according to the observed evolution of the star forming galaxy fraction by \cite{2017AeAdavidzon}. Passive-galaxy contributions are then neglected because they are reported to be very faint in the far-infrared/sub-millimetre \citep[e.g.][]{2021Natur.597..485W}. 
Star-forming galaxies are classified as main sequence (MS) or starbursting (SB) galaxies according to a redshift-evolving fraction described in \cite{bet12}. The SFR of each galaxy is derived using its redshift, $M_*$, and type (MS or SB).

A spectral energy distribution (SED) is attributed to each galaxy based on its redshift, type, total infrared luminosity, $L_{\rm IR}$, and mean radiation field intensity, $\langle U\rangle$. The $L_{\rm IR}$ and $\langle U\rangle$ values are directly derived from the SFR. \\

For CO, the luminosity of the J=1 transition is first computed using the the $L_{\rm IR}$ - $L_{\rm CO(1-0)}$ correlation from \citet{sargent2014}. For SB galaxies, an offset of -0.46\,dex is added to $L_{\rm IR}$. Then, for the rest of the transitions, SLED templates from \citet{bournaud_sled} for MS and from \citet{birkin_sled} for SB are attributed. For MS, contributions to the final SLED of clumpy and diffuse SLEDs are defined by the relation between <U> and CO(5-4)\,/\,CO(2-1) flux ratio from \cite{daddi_2015_COratio}.
 
The two \ci\, fine-structure lines are derived from empirical relations between line luminosity ratios. First the relation between \ci\,(1-0) / $L_{\rm IR}$ and CO(4-3) / $L_{\rm IR}$ is calibrated to derive $L_{\rm [C_I](1-0)}$. Then, the relation between \ci\,(2-1) / \ci\,(1-0) and CO(7-6) / CO(3-4) is calibrated to derive $L_{\rm [C_I](2-1)}$. The calibration of these relations is discussed in B22. \\ 

Finally, two models are implemented to compute the \cii \, luminosity, following two $L_{\rm [C_{II}]}$-SFR relations. The first one is the empirical relation calibrated in the local universe from \cite{2014A&A...568A..62D}, hereafter DL14. The second one is from \cite{2018A&A...609A.130L}, hereafter L18, obtained via a semi-analytical model, and containing a weak redshift dependence. Although the exact $L_{\rm [C_{II}]}$-SFR relation remains debated at high redshifts, the recent observational results of the ALPINE survey \citep{alpines} shows that high-redshift galaxies tend to follow the local $L_{\rm [C_{II}]}$-SFR relation. For this reason, we take \cii\, following the DL14 relation as our fiducial model. 

\subsection{Cube making \label{making}}

From the SIDES catalogue, we create spatial-spectral data cubes for each component: continuum (CIB), CO, \ci\, and \cii \,. The adopted pixel size is 5\,arcseconds and the frequency axis ranges from 125 to 305\,GHz with an absolute spectral resolution of 1\,GHz. 
Voxels (the spatial-spectral elements of a cube) are populated with the corresponding intensities. Lastly, the beam smearing is applied to each frequency channel by convolution in Fourier space with a Gaussian beam kernel and periodic boundary conditions. The Gaussian beam profile in one frequency channel is 
\begin{equation}
    G_{beam} = \frac{1}{2 \pi \sigma^2}\mathrm{e}^{ \frac{-r^2}{2\sigma^2}} \label{2d_beam}\, ,
\end{equation}
where $\sigma$ is the standard deviation of the Gaussian beam profile linked to the frequency $\nu$ of the channel by
\begin{equation}
\sigma = \frac{FWHM}{2 \sqrt{2 ln(2)}} = \frac{1}{2 \sqrt{2 ln(2)}} \frac{1.22 c}{\nu D}\, , \label{sigma_eq}
\end{equation}
with FWHM is the full with at half maximum of the instrument beam, c is the celerity of light and $D=12$\,m is the diameter of the APEX telescope where CONCERTO is installed. All the components, CIB, CO, \ci\,, and \cii,,  are summed together to create a foreground-contaminated cube. This cube represents the mock observation of the sky as observed by CONCERTO. The aim of the deconfusion is to retrieve the \cii\, signal embedded in the foreground-contaminated data. We deal with the two types of foreground one by one, starting with the continuum foreground. 

\section{\label{continuum removal} Continuum removal}

The CIB consists of a smooth, frequency-coherent baseline in the mock electromagnetic spectra, resulting from the sum of dust continuum emission from galaxies along the line of sight (\los \,) plus a part of continuum emission from the neighbouring \los \, due to the beam smearing. The CIB has strong angular fluctuations that are the dominant contribution to the measured power spectrum (in particular, see Fig. 15 of B22). In this section we compare the efficiency of two different methods for continuum removal: the PCA and the asymmetric re-weighted penalized least squares (arPLS).
The analysis was conducted on frequencies ranging from 125 to 305\,GHz. For the \cii\, signal, as the SIDES simulation only covers the redshift range $z\le7$, the analysis was done only for $\nu>$237\,GHz. 
We first assessed the continuum removal on SIDES. SIDES contains both the CIB (mean continuum level) and its spectral fluctuations, which are about 10-20\% of the CIB. But for CONCERTO, even if in theory it would be sensitive to the CIB, the data processing removes the mean levels. We thus also performed the continuum removal after subtracting the CIB at each frequency of the SIDES simulation (i.e. removing the mean level).

\subsection{Principal component analysis \label{PCA}}

Principal component analysis  is a non-parametric method of dimensionality reduction,  based on constructing linear combinations of variables in the dataset.  
Geometrically, these new variables, the so-called principal components, represent the directions of maximum variance in the dataset. The principal components do not have a direct physical interpretation. They are only linked to the variance, not to an underlying physical origin. Most of the variance in the dataset is described by the first principal components, ranked by the amount of variance they carry. Dimensionality reduction corresponds to discarding principal components of lower significance, when projecting the dataset onto this new basis. 
As mentioned above, the CIB produces the strongest fluctuations at all frequencies in CONCERTO. It is therefore likely that its contamination is condensed in the first few principal components and can be accurately removed. We use for this work the PCA implementation from the package scikit-learn \citep{scikit-learn}. We create a matrix, $X$, that contains the collection of spectra from the foreground-contaminated mock cube of Sect.\,\ref{making}. The dimensions of $X$ are $N_{\rm \nu} \times N_{\rm p}$ , where $N_{\rm \nu}$ is the number of frequency channels and $N_{\rm p}$ is the number of pixels. For clarity, we use Greek symbols for the frequency indices and a Latin symbol for the spatial indices ($p$ for pixel). The covariance matrix, $C$, of the spectral data reads
\begin{equation}
    C_{\nu \nu'} = \Tilde{X}_{\nu p}  \Tilde{X}_{p \nu'}\,,
\end{equation}
where $\Tilde{X}$ is defined as
\begin{equation}
 \Tilde{X}_{\nu p} = X_{\nu p} - \mu_p \,,
\end{equation}
and $\mu_{ \rm p}$ is the average in the pixel $p$ over all frequencies. The covariance matrix, $C,$ is square, symmetric, and has dimension $N_{\nu}$. Next, we computed the eigenvector matrix, $E,$ and the block matrix, $\lambda,$ that contains the eigenvalues from the diagonalization of $C$, such that \begin{equation}
     C_{\nu \nu'}=E_{\nu \alpha}^{-1}\lambda_{\alpha \beta} E_{\beta \nu'} \,. 
 \end{equation}

Eigenvectors and eigenvalues are sorted according to the power enclosed in each principal component. We select the $k$ ($k < N_{\nu}$) first principal components and build the block matrix $\lambda'$ with their corresponding eigenvalues. This allows the derivation of what is usually called the projection templates:

\begin{equation}
T_{p \nu}=\Tilde{X}_{\nu' p}^TE^{-1}_{\nu' \alpha}\frac{1}{\sqrt{ {\lambda'}_{\alpha \nu}}} \,,
\end{equation}and the continuum is therefore estimated as
\begin{equation}
    \tilde{X}^{continuum}_{\nu p} = E_{\alpha \nu}^T \sqrt{\lambda}_{\alpha \beta}\,T^T_{p \beta}\, + \mu_p.
\end{equation}Eventually, the line spectrum is given by
\begin{equation}
    X^{lines}_{\nu p} = X_{\nu p} - \Tilde{X}^{continuum}_{\nu p}\,.
\end{equation}

\begin{figure}[htb]
    \centering
    \includegraphics[width=0.45\textwidth]{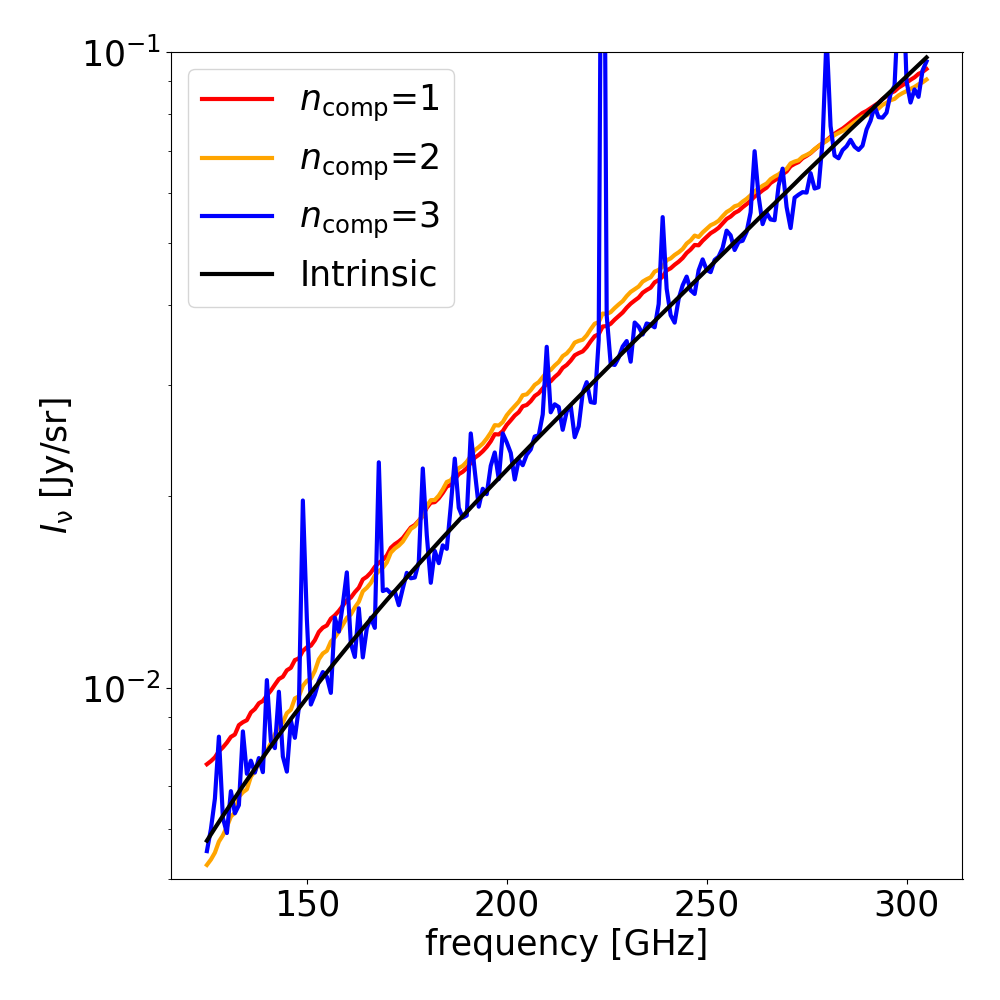}
    \caption{Spectrum of CIB (continuum) emission in one pixel as a function of frequency (black line). The reconstructed spectra obtained with PCA using the first (red), the first two (orange), and the first three (blue) principal components are over-plotted.}  
    \label{pca_spectrum}
\end{figure}

Figure\,\ref{pca_spectrum} shows a representative example of reconstructed continuum spectra with PCA. When using the first two components, the continuum is not well reconstructed, meaning that the continuum features are not totally contained in the first two principal components. When adding the third principal component, the reconstructed spectrum does get closer to the intrinsic continuum spectrum but line emission starts to show up. Hence, it is not possible to reconstruct more accurately the continuum by adding more principal components. 

\begin{figure}[htb]
    \centering
    \includegraphics[width=1\columnwidth]{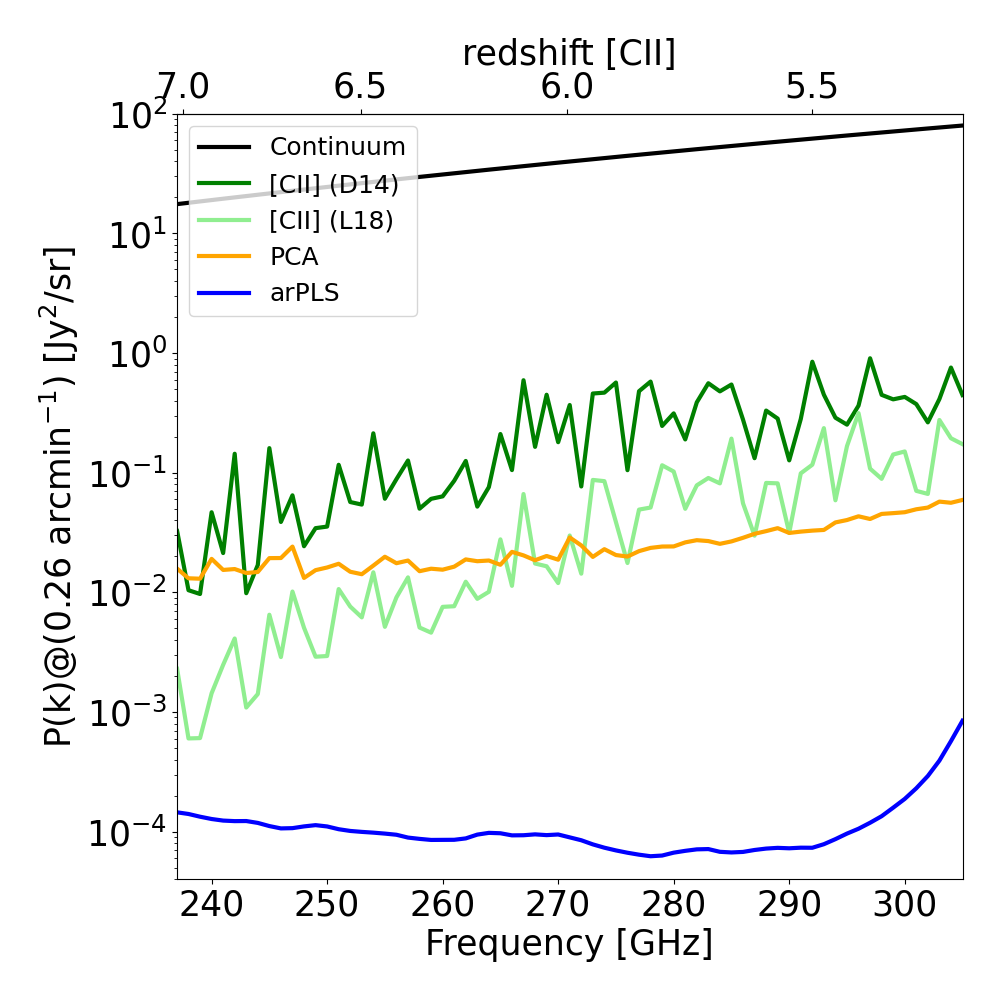}
    \caption{APS amplitude as a function of frequency for the continuum (solid black line), the \cii \,  generated with the DL14 or L18 SFR-$L_{[C_{II}]}$ relations (green and light green lines, respectively), the residual continuum obtained with PCA (Sect.\,\ref{PCA}) using the first two principal components (orange line), and the residual continuum obtained with arPLS (Sect.\,\ref{arpls}, blue line). The power spectrum amplitude is averaged at $k =0.26$ \karcmin\, in frequency channels of 1\,GHz width. }
    \label{cont_residual}
\end{figure}

Figure\,\ref{cont_residual} shows the effect of the reconstructed continuum subtraction on the continuum power spectrum amplitude, using the first two principal components. Around $z$=5.2, the residual continuum is 29 times lower than our fiducial \cii\, model and 7 times lower than the L18 \cii\, model. But the higher the redshift, the lower the contrast. At z=7, the residual continuum is at a similar level than the fiducial  \cii\,  model and 10 times higher than the L18 model. 

When removing the CIB (i.e. the mean level of each frequency channel of the cube), the lines contaminate the continuum estimate starting from the third principal component, as previously. However, removing the CIB decreases the continuum variance, leading to a larger residual continuum power, which is a factor of 11 above the lines signal for $\nu>$237GHz. Hence, PCA does not have the cleaning efficiency required up to redshift 7, and we turned to another method often quoted in the literature that fits the continuum with slowly varying functions.

\begin{figure}[htb]
    \centering
    \includegraphics[width=1\columnwidth]{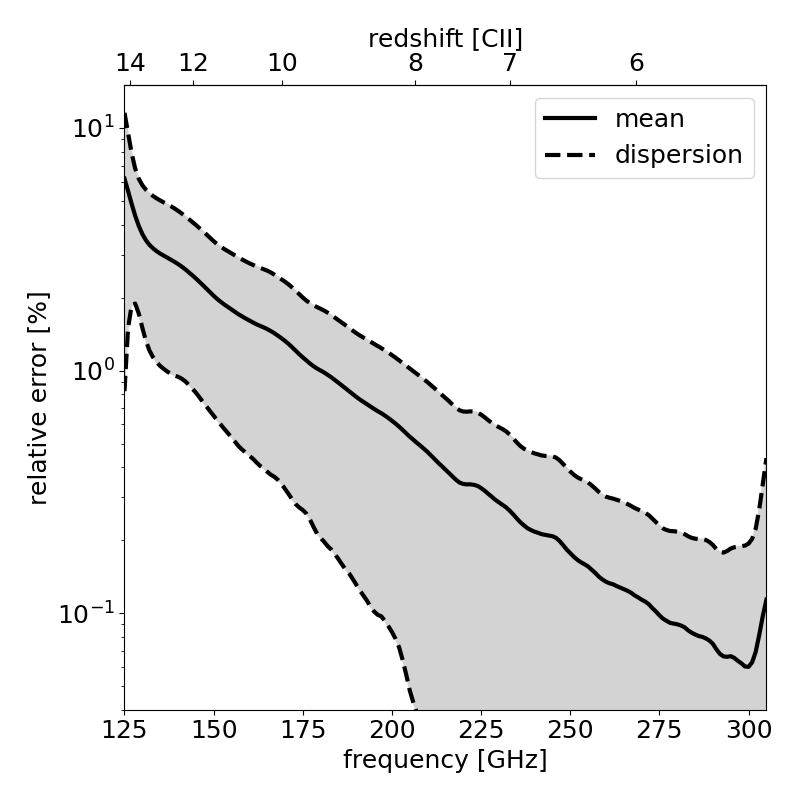}
    \caption{Relative error of arPLS to the `true' continuuum as a function of frequency. The mean is taken over the spectra of all the \los \, of the mock cube.}
    \label{arplsfit}
\end{figure}

\subsection{Advanced baseline estimator: arPLS \label{arpls}}
We test in this section the advanced baseline estimator named arPLS, introduced by \citet{ARPLS} and based a Whittaker-smoothing-based algorithm. Originally developed to serve in a context of Raman spectroscopy, this method was applied to time series in an astronomical context by \citet{arpls_exemple}. Here, we adapted the use of arPLS to spectral vectors. This estimator makes use of the frequency-coherent feature of CIB. It adjusts a slowly varying function to the smooth component of a (electromagnetic) spectrum under the assumption that it varies slowly with frequency. This smooth component (e.g. the black smooth baseline in the spectrum shown in Fig.\ref{pca_spectrum}) is the spectrum of CIB, the sum of all dust continuum emission. Moreover, as the frequency-coherent part of the spectrum is fitted, all other frequency-coherent contamination present as Galactic dust and the CMB are fitted simultaneously. Galaxies are different for each \los\, of the mock cube and can drastically change the slope of the baseline so we need to run the baseline estimator on a pixel-by-pixel basis. The smoothed function is fitted to the data by iteratively minimizing a penalized least squares function: 
\begin{equation}
    S(b)=(y-b)^TW(y-b)+R b^TD^TDb \,,
    \label{pls}
\end{equation}
where $y$ is the spectrum to be fitted, $b$ the smooth baseline to be found, $W$ the block weight matrix with the weights $w$ for each data point, and $D$ the difference matrix. It is a quasi blind method as the only parameter to be set is the regularization parameter $R$ controlling the trade-off between the agreement of $b$ to the data and its smoothness. In this work, we set it to 1. To find the line peak regions, weights are asymmetrically given to data points: 

\begin{equation}
    \label{eq:weights}
    \mathrm{w_i = \left\{ \begin{array}{l} f(y_i-b_i, m, \sigma),\, y_i \geq b_i \\ 1, \,y_i < b_i \end{array} \right. }
,\end{equation}
with
\begin{equation}
    \label{eq:f}
    f(d^{-},m,\sigma)=\frac{1}{1+\mathrm{exp}(2(d-(-m+2 \sigma))/\sigma) }
,\end{equation}
where $d^{-}=y^{-}-b$ represents the distance between the current estimate of $b$ and the data points  that are positioned exclusively below it. The $m$ and $\sigma$ are the mean and dispersion $y^{-}$ of $d$. 

This weighting scheme is designed to be robust to reasonable noise level blurring the baseline. Here, we first want to test the ability of this estimator to retrieve the baseline of the simulated LIM electromagnetic spectra along the CONCERTO frequency band, without any instrument noise. 

Figure\,\ref{arplsfit} shows the relative error between the continuum fitted with arPLS and the intrinsic continuum (from the CIB cube). While the baseline is slightly overestimated at all frequencies, the mean error stays below 0.3$\%$ for $\nu \geq 237\rm GHz$ (\zcii \,$\leq$\,7). 

Figure\,\ref{cont_residual} shows that the power amplitude of continuum is brought below the fiducial \cii\, by a factor of >72 and below the L18 \cii\, by a factor >4, for $z=5.-72$. This allows a clear detection of the \cii\, power even in the fainter \cii \, case of L18. 

For $\nu \leq 237\rm GHz$, the mean relative error and its dispersion become larger as there are more and more CO lines. Around 130\,GHz, at the end of the frequency band, the accuracy decreases even more since there are fewer neighbouring frequency channels to correlate with. Despite this, the fluctuations of the residual continuum are about 27 times lower than the CO power amplitude and permit a further investigation of the CO power spectrum at those frequencies (below 237\,GHz). 

Finally, we tested arPLS after CIB removal. In that case, the residual power continuum has the same amplitude as before (blue line in Fig. \,\ref{cont_residual}) but it is noisier, with a standard deviation of 10 times larger. In any case, the residual contamination is still lower than the L18 \cii\, power spectrum above 237\,GHz.

We conclude that this method is appropriate for removing the continuum emission from galaxies for CONCERTO, incidentally better than a standard PCA, and is worth being investigated with noise realizations. In what follows, we detail the interloper component separation method. 

\section{\label{sect:masking} Removing the contamination of line interlopers}

Once the continuum power amplitude is successfully lowered, the main source of contamination becomes the line interlopers, that is to say, the emission lines redshifted in the same frequency band as \cii \,. Dominant interlopers are the CO rotational ladder and two \ci\, fine structure lines from foreground galaxies at $z<3$. The [$\rm O_I$] 145 $\rm \mu m$, [$\rm N_{II}$] 121.9 $\rm \mu m$ and $\rm N_{II}$ 205.2 $\rm \mu m$ lines are also present in the CONCERTO frequency range but their intensities are negligible with respect to the \cii \, intensity \citep{silva_masking_cstlinelfux}. Consequently, only CO and \ci\, interlopers are considered. 

The LIM survey exploits the confusion regime from the far-IR to (sub-)millimetre, where the beam smearing is significant. Because of the confusion regime and together with the expected instrument noise level for CONCERTO \citep{concerto_2020} we do not expect to detect lines from individual foreground galaxies. It becomes thus impossible to blindly mask the brightest pixels using a fix intensity threshold. Therefore, to select interloper-dominated voxels, we need to rely on external catalogues of the COSMOS field \citep{ilbert13_cosmos,L16,cosmos_candels_2017,cosmos2020} providing accurate positions and redshifts of the foreground sources. In this section we present the masking strategy based on the ancillary catalogue of detected sources in the COSMOS field. We first justify our choice of using an external catalogue in stellar mass $M_*$ and its use as a CO proxy to select foreground sources. We then detail the masks we build from this ancillary catalogue.

\subsection{Stellar mass as a CO power proxy \label{selection_sources}}

\begin{figure*}[htp]
\includegraphics[width=\textwidth]{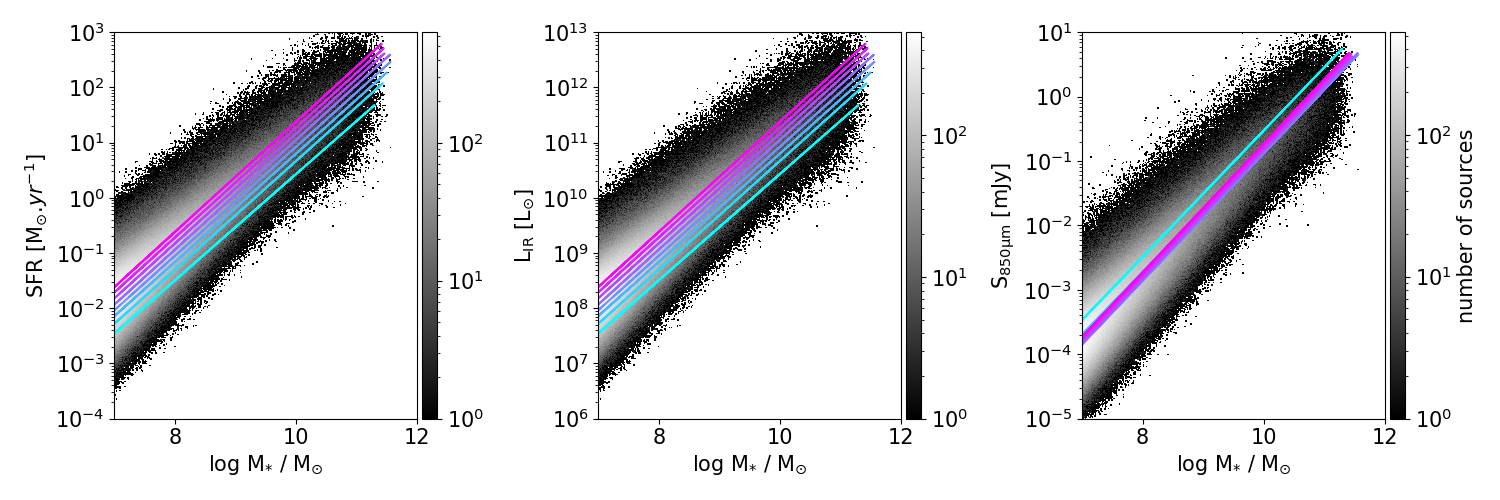}
\caption{Stellar mass distribution of sources as a function of different CO proxies available in SIDES.
  The stellar mass distribution of SIDES sources is shown in grey scale as a function of different proxies up to $z<3.5$ (from left to right): $M_{*}$ - SFR, $M_{*}$ - $L_{IR}$, and $M_{*}$ - $S_{850}$. The best fits for each relation in each redshift bin (given in Table\,\ref{mstar_vs_key_tab}) are over plotted, from cyan for the first bin (0< z < 0.35) to pink for the last bin (2.75 < z < 3.5).  \label{mstar_vs_keys} }
\end{figure*}

\begin{table*}
 \centering
 \caption[]{Equivalent depth in SFR, $L_{\rm IR}$, and $S_{\rm 850 \mu m}$ to the $M^{\rm L16}_{*}$ depth of the COSMOS completeness relation, obtained with the best fits from Fig.\,\ref{mstar_vs_keys}. \label{mstar_vs_key_tab} }

\tiny{
\begin{tabular}{c c | c c c }
\hline
\hline
\noalign{\smallskip}
Redshift & log $M^{\rm L16}_* / \rm M_{\odot}$ & SFR [$\rm M_{\odot}.yr^{-1}$] & $L_{\rm IR}$ [$\rm L_{\odot}$] & $S_{\rm 850 \mu m}$ [$\rm mJy$] \\
\noalign{\smallskip}

\hline
0-0.35 & 8.1 & $(4.0 \pm 1.3) \times 10^{-2}$ & $(4.0 \pm 1.3) \times 10^{8}$ & $(4.1 \pm 1.6) \times 10^{-3}$ \\

0.35-0.65 & 8.7 &  $(2.5 \pm 0.8) \times 10^{-1}$ & $(2.5 \pm 0.8) \times 10^{9}$ & $(9.1 \pm 3.2) \times 10^{-3}$ \\

0.65-0.95 & 9.1  & $(8.3 \pm 2.7) \times 10^{-1}$ & $(8.3 \pm 2.7) \times 10^{9}$ & $(1.8 \pm 0.6) \times 10^{-2}$ \\

0.95-1.3 & 9.3 & $1.7  \pm 0.6 $ & $(1.7 \pm 0.6) \times 10^{10}$ & $(2.7 \pm 0.9) \times 10^{-2}$ \\

1.3-1.75 & 9.7 & $5.7  \pm 1.9 $ & $(5.7 \pm 1.9) \times 10^{10}$ & $(6.9 \pm 2.4) \times 10^{-2}$ \\

1.75-2.25 & 9.9 & $(1.2 \pm 0.4) \times 10^{1}$ & $(1.2 \pm 0.4) \times 10^{11}$ & $(1.2 \pm 0.4) \times 10^{-1}$ \\
2.25-2.75 & 10.0 & $(1.9 \pm 0.6) \times 10^{1}$ & $(1.9 \pm 0.6) \times 10^{11}$ & $(1.7 \pm 0.6) \times 10^{-1}$ \\

2.75-3.5 & 10.1 & $(3.0 \pm 1.0) \times 10^{1}$ & $(3.0 \pm 1.0) \times 10^{11}$ & $(2.3 \pm 0.8) \times 10^{-1}$ \\

\hline
\end{tabular}}

\end{table*}

For typical redshifts of foreground sources in COSMOS ($z\sim\,1-2$), the selection wavelengths of the COSMOS catalogues trace the stellar emission, that traces the stellar mass, rather than the ultraviolet, that traces the unobscured SFR. Therefore, the stellar mass is usually more complete in those catalogues than the SFR, from which moreover the obscured contribution is often missing.

Figure\,\ref{mstar_vs_keys} shows the distribution in $M_{*}$ for galaxies $z \leq 3.5$ as a function of other quantities that can also serve as CO proxy available in SIDES: SFR, $L_{\rm IR}$ and flux at 850\,µm $S_{850}$. From these distributions, we obtain the best fit relation for objects in each redshift bin of the completeness relation, taken from \citet[hereafter L16]{L16} and reported in Table\,\ref{mstar_vs_key_tab}. The completeness relation of the COSMOS catalogues describes, as a function of redshift, the stellar mass down to which 90\,$\rm \%$ of sources are detected. Then from the best fits, we get equivalent depths to the stellar mass depths $M^{L16}_{*}$.
Results are listed in Table\,\ref{mstar_vs_key_tab}. One should reach these depths to get a catalogue as complete in these proxies as in stellar mass. In comparison, $L_{\rm IR}$ values inferred from the Multi-Band Imaging Photometer (MIPS)/\textit{Spitzer} at 24\,$\mu m$ \citep{2005ApJ...632..169L, 2008A&A...479...83B} are $1 \times 10^{10} \rm L_{\odot}$ at $z = 0.3$, $3 \times 10^{10} \rm L_{\odot}$ at $z =0.6$, and $1 \times 10^{11} \rm L_{\odot}$ at $z=1$. It is one order of magnitude above the $L_{\rm IR}$ depth needed. About SFR, no survey has such a completeness in both ultraviolet and IR to compute the total SFR. \cite{Geach_2016} report the Submillimetre Common-User Bolometer Array 2 (SCUBA-2) 850\,$\rm \mu$m confusion limit to be 0.8\,mJy/beam with a completeness below 10\,$\rm \%$ at this flux. Completeness is close to 100\,$\rm \%$ above 5\,mJy/beam, which is largely above (between 1 and 3 orders of magnitude) the $S_{850}$ requested depth.

To further motivate this masking strategy using SIDES, we look at the contribution of each stellar mass population to the CO power spectrum. Especially, we want to make sure that (1) most of the CO power is produced by sources above the L16 completeness relation, so that they can actually be detected and masked in real data, and (2) that CO power is produced by a reasonable amount of sources, so that masking them would not cost too much in survey area. 

\begin{figure}
\centering
\includegraphics[width=\columnwidth]{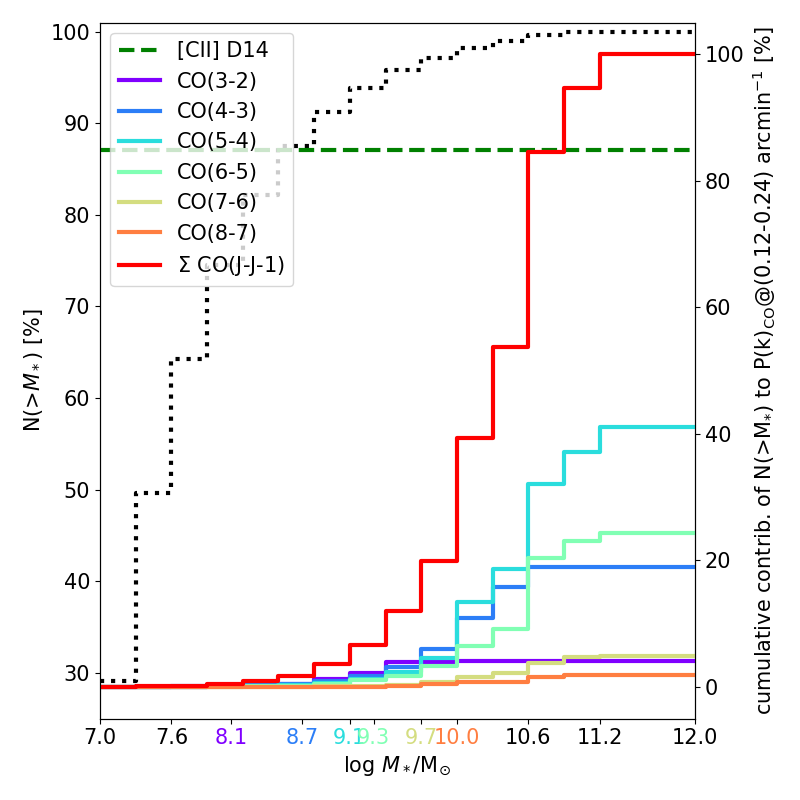}
\caption{ Contribution to the total CO APS at $z=5.2$ of each stellar mass bin and CO transition. The left axis and the dotted black curve show the  cumulative number count of CO emitters per bin of stellar mass (in percent) redshifted in the frequency channel 305\,$\pm$0.5\,GHz (\zcii \,=\,5.2). The right axis shows the cumulative contribution (in percent) to the total CO power amplitude. Coloured curves correspond to each CO transition, and the red curve corresponds to the sum of all CO transitions. The 90\% stellar mass completeness in COSMOS2015 for each CO line is indicated with the same colour code on the x-axis. For comparison, the \cii\, level (in percent) is represented with the dashed green line. Power amplitudes are averaged for $0.12 < k < 0.24$\,\karcmin\,.\label{fig:mstar_bin_1}}
\end{figure}

Figure \,\ref{fig:mstar_bin_1} shows the cumulative contribution to the CO power amplitude per increasing logarithmic step of stellar mass at \zcii\,=5.2. By looking at the cumulative contribution rather than the non-cumulative contribution, we take into account for each stellar mass population (i.e. at each step) both its auto-correlation and its cross-correlation with the less massive populations. The cross-correlation contribution is actually important: the non-cumulative contributions (i.e. the auto-correlations alone) of all stellar mass populations represent 60\% of the total CO power. The remaining 40\% is due to cross-correlations between the different stellar masses. Figure\,\ref{fig:mstar_bin_1} shows that the most massive galaxies, by their auto and cross correlations, contribute the most to the CO power. This is also the case at higher redshifts: at \zcii\,=\,6.5 and \zcii\,=7. The correspondence between \cii\, redshifts and interloper redshifts is given in Table\,\ref{redshift}. 

For the three redshifts, galaxies more massive than $ \geq 10^{8.1} \rm M_{\odot}$ represent only 22-25\,$\rm \%$ of the total object number but produce most of the CO interloping power (99.7\,$\%$). On the other hand, 75-77$\%$ of objects less massive than $10^{8.1} \rm M_{\odot}$ are responsible for less than 0.5$\%$ of the total CO power. For example, we can highlight that the galaxy population with $10^{10.6} \rm M_{\odot}<M_*\leq 10^{10.9} \rm M_{\odot}$ is responsible for 24-31$\%$ of the CO power while they only represent 0.5-0.8$\%$ of all the sources. Hence, the  massive sources above $10^{8.1}$\,M$_\odot$, despite being the less numerous, strongly dominate the contribution to the CO power and belong to the massive class of known objects. This makes masking an efficient method for decreasing the CO contamination by removing a few sources without catastrophically masking the survey area. Hence, the stellar mass $M_*$ is a reliable CO proxy for the masking purpose. \\ 

To make a realistic masking simulation, we masked only sources in the SIDES catalogue that are above the COSMOS completeness relation. We kept the completeness taken from L16, which is deeper than that from \cite{ilbert13_cosmos} or \cite{cosmos_candels_2017}. Although the completeness reported in \cite{cosmos2020} is deeper, it only corresponds to a 70\% completeness fraction, so the L16 relation represents the best compromise between completeness fraction and depth. 

\subsection{Effective masking \label{effective_masking}}

\begin{figure*}
    \centering
    \includegraphics[width=0.8\textwidth]{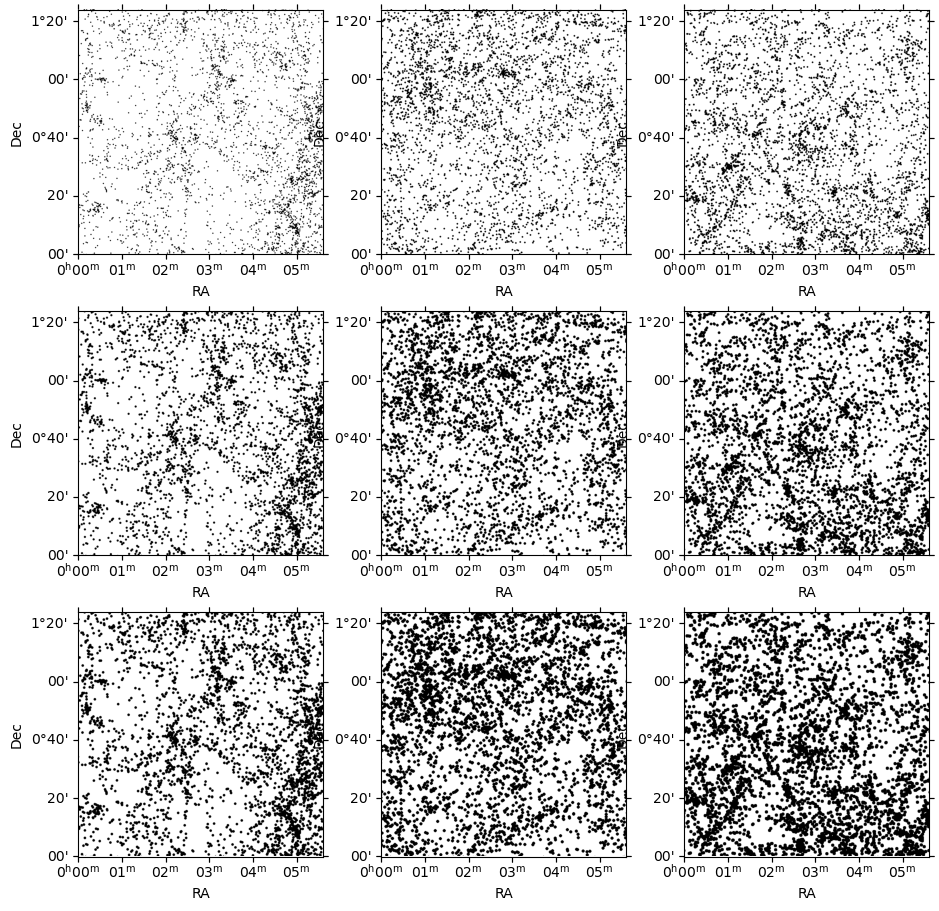}
    \caption{Masks obtained following the strategy detailed in Sect.\,\ref{sect:masking} using (from left to right) a masking radius, $r_{\sigma}$, of 1.5\,$\sigma$, 2.5\,$\sigma$, and 3\,$\sigma$, respectively. From the top to the bottom, the rows correspond to  305\,GHz (\zcii \,=\,5.2), 253\,GHz (\zcii \,=\,6.5), and 237\,GHz (\zcii \,=\,7), respectively. Masked data appear in black.}
    \label{masks}
\end{figure*}

\begin{table}
\caption[]{Emission redshift for the lines at the three observed frequencies. \label{redshift}}
\tiny{
\begin{tabular}{llccc}
\hline
\hline
\noalign{\smallskip}
Line & $\rm \nu_{ref}$\,(GHz) & $\rm z(305\,GHz)$ & $\rm z(253\,GHz)$ & $\rm z(237\,GHz)$  \\
\hline
\noalign{\smallskip}
CO(3-2) & 345.81  & 0.13 & 0.37 & 0.46 \\
\noalign{\smallskip}
CO(4-3) & 461.08  & 0.51 & 0.82 & 0.95 \\
\noalign{\smallskip}
CO(5-4) & 576.36  & 0.89 & 1.28 & 1.43 \\
\noalign{\smallskip}
CO(6-5) & 691.63  & 1.27 & 1.73 & 1.92 \\
\noalign{\smallskip}
CO(7-6) & 806.9  & 1.65 & 2.19 & 2.4 \\
\noalign{\smallskip}
CO(8-7) & 922.17  & 2.02 & 2.64 & 2.89 \\
\noalign{\smallskip}
\ci\,(1-0) & 492.16  & 0.61 & 0.95 & 1.08 \\
\noalign{\smallskip}
\ci\,(2-1) & 809.34  & 1.65 & 2.20 & 2.41 \\
\noalign{\smallskip}
\cii \, & 1900.54  & 5.23 & 6.51 & 7.02 \\
\noalign{\smallskip}
\hline
\end{tabular}}
\end{table} 

External catalogues provide the position of the sources that we want to mask. We then needed to define the extent of the region to mask around these positions. We based our choice on a beam width criterion. Voxels were masked around the location of the selected sources (and thus around CO or \ci\, expected emission) within a radius, $r_{\sigma}$, expressed in units of the Gaussian beam width parameter, $\sigma$ (Eq.\,\ref{sigma_eq}). This beam varies with the observed frequency and so does the exact masked area. Table\,\ref{masked_fraction} lists the fraction of masked intensity per source and the masked fraction of the maps for masks built with different values of $r_{\sigma}$. Figure\,\ref{masks} displays examples of these masks.

Estimating the APS on masked data is not straightforward because the mask induces aliasing. Several estimators have been developed to overcome this difficulty in the context of the CMB measurements. In this work, we used the P of K EstimatoR \citep[\poker;][]{poker}, which was adapted to the case of sky patches of a few square degrees, with arbitrary high angular resolution and complex masks \cite{2011A&A...536A..18P,2014A&A...571A..30P}. A detailed discussion on the performance and validation of \poker\, in the context of this work and the derivation of error bars on APS are presented in Appendix\,\ref{poker_appendix}.

\begin{table}
\caption[]{Masked intensity per source and survey area loss at the three observed frequencies, corresponding to \zcii \,=\,5.2, 6.5, and 7 for the masks built with different $\rm r_{\sigma}$. \label{masked_fraction}}
\centering
\tiny{
\begin{tabular}{p{0.1\linewidth}|p{0.15\linewidth}p{0.15\linewidth}p{0.15\linewidth}p{0.15\linewidth}}
\hline
\hline
\noalign{\smallskip}
 & Masked  & Masked  & Masked  & Masked  \\
$r_{\sigma}$ & intensity   & fraction [$\%$] &  fraction [$\%$] &  fraction [$\%$] \\
             &  per source & @305\,GHz       & @253\,GHz         & @237\,GHz  \\
\noalign{\smallskip}
\hline
\noalign{\smallskip}
1.5 & 95\,$\%$ & 7 & 11 & 14  \\
2.0 & 99\,$\%$ & 11 & 15 & 22  \\
2.5 & >99\,$\%$ & 16 & 22 & 29  \\
3.0 & >99\,$\%$ & 22 & 29 & 38  \\
\noalign{\smallskip}
\hline
\end{tabular}}
\end{table}

\section{Interloper separation at $z_{[C{II}]}$\,=\,5.2, 6.5, and 7 \label{sect:masking_result}}
\begin{table}
\centering
\caption[]{ Power amplitude (in $\rm Jy^2$/sr) averaged between $k$\,=\,0.12 and 0.24 arcmin$^{-1}$ of the different components, before and after the corresponding component separation method is applied for foreground removal. Error bars on the CO, \ci\, and \cii\, amplitudes were computed using the dispersion measured on 54 independent fields (with \rsigma\,=\,3.0). Details are given in Appendix\,\ref{poker_appendix}. \label{tab:pk_values}}
\tiny
\begin{tabular}{l*{3}{c}}
\hline
\hline
\noalign{\smallskip}
Frequency & 305\,GHz & 253\,GHz & 237\,GHz  \\
\hline
\noalign{\smallskip}
& \multicolumn{3}{c}{ Intrinsic amplitudes } \\
\noalign{\smallskip}
\hline
\noalign{\smallskip}
$ \rm P(k)_{CIB} $ & $1.1 \times 10^{2}$  & $3.6 \times 10^{1}$  & $2.4 \times 10^{1}$  \\
\noalign{\smallskip}
$ \rm P(k)_{CO} $ & $6.2 \times 10^{-1}$  & $4.1 \times 10^{-1}$  & $5.1 \times 10^{-1}$  \\
\noalign{\smallskip}
$ \rm P(k)_{[CI]} $ & $2.9 \times 10^{-1}$  & $1.1 \times 10^{-1}$  & $1.7 \times 10^{-1}$  \\
\noalign{\smallskip}
$ \rm P(k)_{[CII]} $ & $5.3 \times 10^{-1}$  & $6.0 \times 10^{-2}$  & $3.4 \times 10^{-2}$  \\
\noalign{\smallskip}
\hline
\noalign{\smallskip}
& \multicolumn{3}{c}{CIB residual amplitudes} \\
\noalign{\smallskip}
\hline
\noalign{\smallskip}
 $\rm PCA$  & $1.5 \times 10^{-1}$ & $2.8 \times 10^{-2}$ & $2.7 \times 10^{-2}$ \\
\noalign{\smallskip}
 arPLS  & $1.1 \times 10^{-3}$ & $1.5 \times 10^{-4}$ & $1.9 \times 10^{-4}$ \\
\noalign{\smallskip}
\hline
\noalign{\smallskip}
& \multicolumn{3}{c}{ \rsigma\,=\,3.0 masked amplitudes} \\
\noalign{\smallskip}
\hline
\noalign{\smallskip}
$ \rm P(k)_{CO+[C_I]} $ & $(8.7 \pm  2.3) \times 10^{-3}$ & $(1.1 \pm  0.4) \times 10^{-2}$ & $(1.7 \pm  0.6) \times 10^{-2}$ \\
\noalign{\smallskip}
$ \rm P(k)_{CO\,bright} $ & $(9.2 \pm  2.4) \times 10^{-5}$ & $(2.0 \pm  0.7) \times 10^{-5}$ & $(3.3 \pm  1.2) \times 10^{-5}$ \\
\noalign{\smallskip}
$ \rm P(k)_{CO\,faint} $ & $(6.6 \pm  1.7) \times 10^{-3}$ & $(7.7 \pm  2.5) \times 10^{-3}$ & $(1.2 \pm  0.4) \times 10^{-2}$ \\
\noalign{\smallskip}
$ \rm P(k)_{[CI]\,bright} $ & $(5.3 \pm  1.4) \times 10^{-6}$ & $(3.7 \pm  1.2) \times 10^{-6}$ & $(3.1 \pm  1.1) \times 10^{-6}$ \\
\noalign{\smallskip}
$ \rm P(k)_{[C_I]\,faint} $ & $(1.9 \pm  0.5) \times 10^{-3}$ & $(2.4 \pm  0.8) \times 10^{-3}$ & $(3.6 \pm  1.3) \times 10^{-3}$ \\
\noalign{\smallskip}
$ \rm P(k)_{[C_{II}]} $ & $(5.3 \pm  1.4) \times 10^{-1}$ & $(7.1 \pm  2.3) \times 10^{-2}$ & $(3.0 \pm  1.0) \times 10^{-2}$ \\
\noalign{\smallskip}
\hline
\end{tabular}
\end{table}

\begin{figure*}
    \centering
    \includegraphics[height=0.35\textwidth]{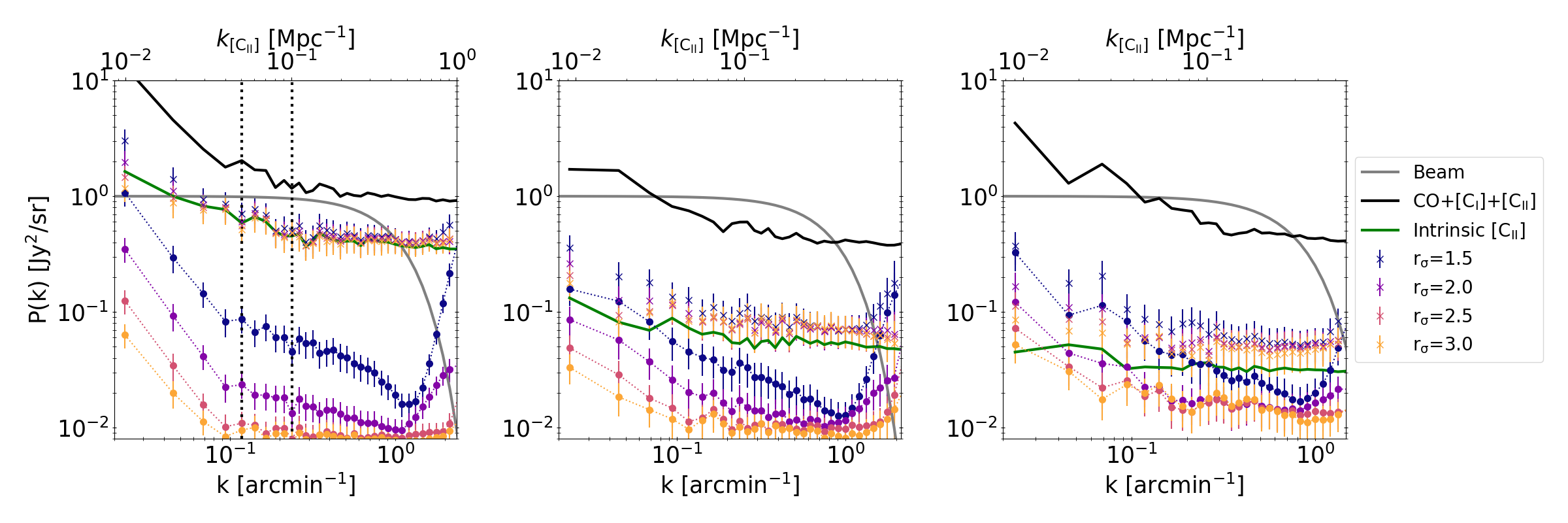}
    \caption{ APS as a function of spatial frequency, $k$. The solid black line shows the APS of the interloper-contaminated map (i.e. CO+\ci\,+\cii\,) with no mask. The solid green line shows the intrinsic \cii\, APS (from the unmasked contamination-free \cii\, map). Coloured crosses show the \cii\, APS estimates, obtained from masking the interloper-contaminated map with the different masks from $r_{\sigma}$\,=\,1.5 to 3.0. Coloured points linked by dotted lines show the residual interlopers' APS measured on the masked interloper map (i.e. CO+\ci\, after masking) at (from left to right) $z$=5.2, 6.5, and 7. The beam APS at the corresponding observed frequency is shown in grey.
    Vertical lines highlight the range $k$\,=\,0.12-0.24\,\karcmin\,. }
    \label{fig:masking_results_three_z}
\end{figure*}
   
In this section we assess the efficiency of masking at retrieving the underlying \ciidl\, APS in the interloper-contaminated mock LIM data (see Sect.\ref{effective_masking}).
As mentioned above, the correction of the mask impact is taken care of by \poker\,, so we focus here on the residual contribution of CO and \ci\, left after masking on the COSMOS field. When using an APS measurement to derive cosmological parameters, one must also account for the FtF variance, namely the overall uncertainty on the relative contribution of CO and \ci\, to \cii, in different parts of the sky. This particular point is addressed in Sect.\,\ref{sect:ftf_var}, and we focus here on the measurement on COSMOS only. 

In the following, power amplitudes and power ratios are given for the same $k$ range (as in Table\,\ref{tab:pk_values}), where the clustering dominates. This is because the clustering part of the APS should be easier to interpret than the shot noise, which is dominated by the few strongest sources just below the detection threshold (G22), thus lowering the statistical information.  \\

Results are presented in Fig.\,\ref{fig:masking_results_three_z} at \zcii \,=\,5.2, 6.5, and 7 for different \rsigma\,. With \rsigma\,=3, the interloper power amplitude is smaller than $2 \times 10^{-2}$\,Jy$^2$/sr for all three redshifts, that is, it is reduced by a factor of $(100 \pm 30)$, $(50 \pm 16),$ and $(42 \pm 15)$, for each redshift, respectively.

For \cii\,, our fiducial model that follows the local DL14 ${\rm L}_{\rm [C_{II}]}$--SFR relation predicts a drop in power amplitude by a factor of 16 between $z=5.2$ and 7. In this framework, masking with \rsigma\,=\,2 at \zcii \,=\,5.2 is enough to obtain a contrast between \cii\, and residual interlopers of \cii\,/(residual CO+\ci\,)\,=\,$(30 \pm 16)$ (i.e. 3\% of residual contamination). The contrast goes up to $(62\pm 32)$ with \rsigma\,=\,3. At \zcii \,=\,6.5, the \cii\, APS is overestimated. Even masking with \rsigma\,=\,3 does not achieved a better power ratio than \cii\,/(residual CO+\ci\,)=\,$(5.5 \pm 3.6)$ (i.e. 18\% of residual contamination after masking). Finally, at \zcii \,=\,7, the contrast between \cii\, and residual interlopers is weak due to the low \cii\, APS, causing \cii\,/(residual CO+\ci\,)=\,$2.0 \pm 1.4$ (i.e. 50\,$\rm \%$ of residual contamination after masking) with \rsigma\,=\,3. 

These results raise several questions that we discuss in the next section regarding: how well the \cii \, APS is reconstructed, especially at redshift 5.2; the cause of the residual power spectra at \zcii \,=\,6.5 and \zcii \,=\,7; and whether increasing \rsigma\ to >3 helps in reducing the residual contamination. To answer these questions, we study separately the masking of \cii\,, free from contamination, the masking of the bright (i.e. above the mass threshold) sources and the masking of the faint (i.e. below the mass threshold) sources. 

\subsection{Interloper masking \label{sect:result_masking}}

\begin{figure*}
\centering
\hfill
\includegraphics[width=\textwidth]{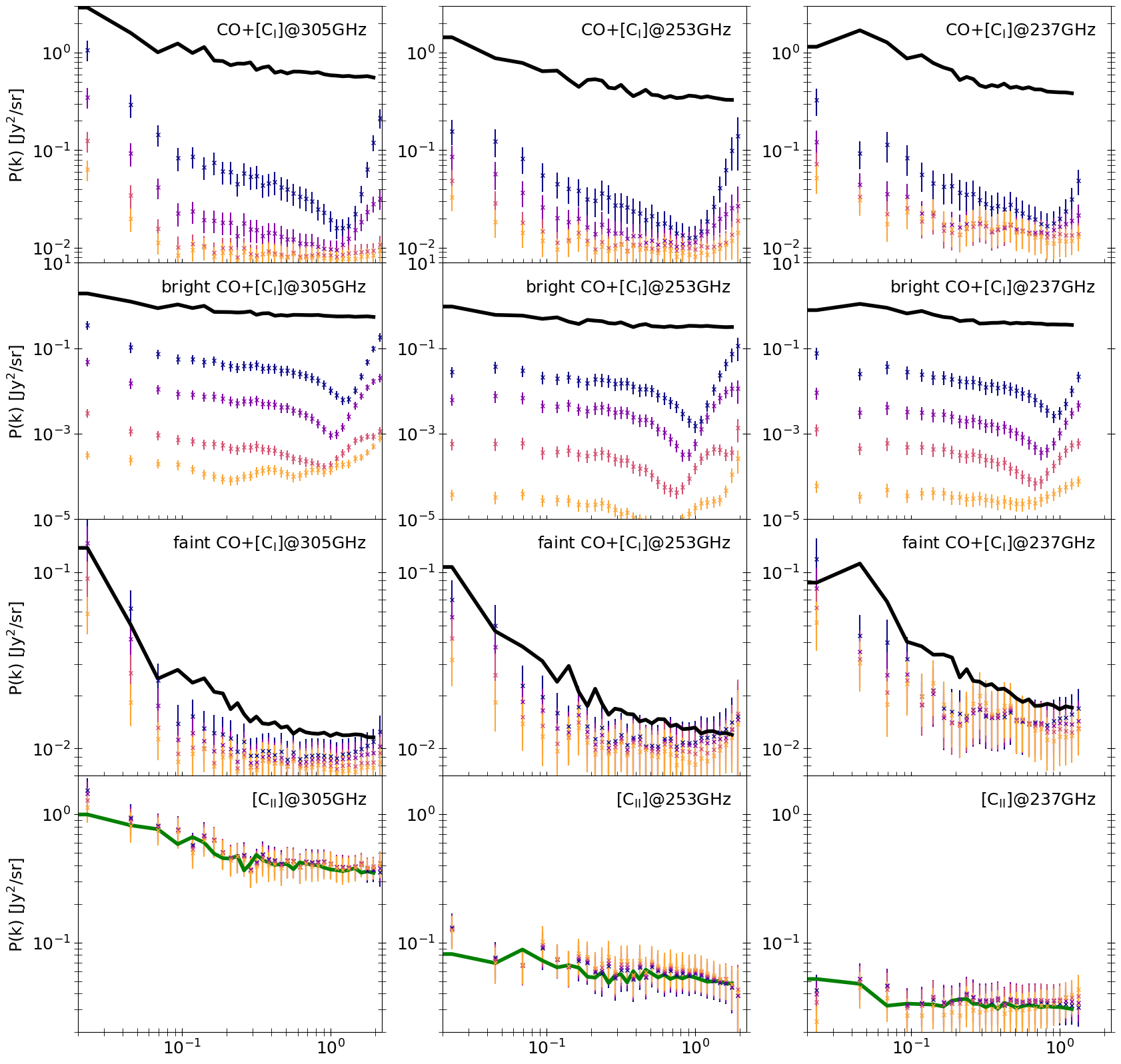}
\caption{APS (in $\rm Jy^{2}/sr$) as a function of spatial frequency $k$ (in \karcmin\,) for the three observed frequencies (from left to right): 305, 253, and 237\,GHz. This is shown (from top to bottom) for each component separately: all the interlopers (CO+\ci\,), the bright interlopers (i.e. above the mass threshold), the faint interlopers, and \cii \,. The intrinsic power spectra of each (unmasked) component are shown with solid black lines, and with a solid green line for \cii \,. The power spectra obtained from the maps that are masked using $r_{\rm \sigma}$= 1.5 to 3.0 are shown with coloured points (from blue to yellow, respectively).} 
\label{fig:species}
\end{figure*}
    
The power spectra from the interloper-contaminated maps measured in the previous section are the sum of the power spectra of the three components (\cii\,, the bright interlopers and the faint interlopers), all affected by the mask. In Fig.\,\ref{fig:species}, we show what we obtain when using different masks on each component separately, for the three redshifts. Besides, Table\,\ref{tab:pk_values} compiles the power amplitudes of each component.

We first focus on the APS behaviour when masking the faint and bright sources, as displayed in the second and third rows of Fig.\,\ref{fig:species}. With \rsigma\,$\leq 1.5$, the residual APS shows an overshoot at high $k$. This is due to the residuals on the edges of the bright sources (`wings') that are still dominating the contribution to the APS, when the mask is too narrow compared to the instrument beam.
When masking at \rsigma\,$\geq 2$, the residual contribution of bright sources drops below $10^{-3}$\,Jy$^2$/sr and becomes negligible compared to that of the faint sources that stays about $0.8\times 10^{-2}$ Jy$^2$/sr. The final power ratio (with \rsigma\,=\,3) between faint and bright sources are (faint CO+\ci\,)\,/ (bright CO+\ci\,)\,=\,$(91 \pm 47)$, $(470 \pm 300),$ and $(480 \pm 330)$ at \zcii\,=\,5.2, 6.5, and 7, respectively. Interestingly, the third row shows how the APS of faint sources is also affected by the mask, contrary to what was expected in B22. Using SIDES, we indeed see that \rsigma\,=\,3 masking lowers the faint power amplitude by a factor of $(1.9 \pm 0.7)-(2.5 \pm 0.6)$ at z=7-5.2. An explanation for this could be that a fraction of faint sources is clustered around bright sources of the same redshift. 
Yet, if masking removes bright sources efficiently, it has only a moderate effect on the faint sources. Indeed, increasing $r_\sigma$ from 1.5 to 3.0 increases the masked area by a factor of 3.1, 2.6, and 2.7 at z=5.2, 6.5, and 7, but reduces the faint signal by only an extra 14\%, 10\%, and 1\%, respectively, while in comparison the bright signal is reduced by over a thousand. As a result, enlarging the masked area around bright sources does not effectively remove more contamination from the faint sources and increases rapidly the loss of survey area.  

We also applied the masking procedure to \ci\, and CO maps separately. \cite{sun_masking} (hereafter S18) studied the masking of CO emitters having a magnitude $m_K^{AB} \la 22$, which translates to a stellar mass cut evolving with redshift \citep[see Fig.\,7 in][]{sun_masking}. This lowers their CO power amplitude by a factor $\geq 100$ at $z=6.5$ and $k=(3-6)\,\times\,10^{-2}\, \rm h/Mpc$, for a 8\,$\%$ loss of survey area. With SIDES, at comparable redshift and spatial frequencies (corresponding to $k=0.12-0.24$\,\karcmin\, at $z=6.5$), masking with this prescription lowers the CO power amplitude by only a factor of 20 for a 22$\%$ loss of survey area. This indicates that the difference in the CO power model between S18 and SIDES increases the power faint sources in our mock data. Faint sources left after masking contributes at least 5 times more in SIDES than in S18. Moreover, taking into account the beam smearing when masking increases significantly the masked fraction of the surveyed area. With our strategy of masking all detected sources that have a magnitude $m_K^{AB} < 24$ in \citep{L16}, the CO power amplitude is lowered by a factor of 50, for a 29\% loss of survey area (with \rsigma\,=3). Lastly, CO is still the dominant interloper after masking: the interlopers' power ratio is (residual CO\,/ residual\,\ci\,)\,=\,$(3.2 \pm 2.1)$, on average for the three redshifts studied (\zcii\,=5.2, 6.5 and 7). 

With the understanding of how masking affects the interlopers, we show in the last row of Fig.\,\ref{fig:species} the \cii \, APS reconstructed from the masked \cii\, maps of SIDES-Bolshoi. The reconstructed \cii\, power amplitude is consistent with the intrinsic one within $1\rm \sigma$. However, Appendix\,\ref{poker_appendix} details how the measured \cii\, APS (using the mask \rsigma\,=\, 3) are biased low, on average by 1\%, 7\%, and 8\,$\rm \%$ at \zcii\,=\,5.2, 6.5, and 7, respectively,  for 54 Uchuu fields. 

In conclusion, the masking strategy with \rsigma\,=\,3 is able to nicely reconstruct the \cii \, power spectrum from the SIDES-Bolshoi realization at \zcii \,=\,5.2, reaching a power ratio \cii \,/\,(residual CO)\,=\,$(80 \pm 42)$ and \cii \,/\,(residual \ci\,)=\,$280 \pm 150$, with a bias on the \cii\,APS of 1\,$\%$ due to the mask. At \zcii\,=6.5, the power ratios are \cii \,/(residual CO)\,=\,($7.8 \pm 5.1$) and \cii \,/\,(residual \ci\,)=\,($25 \pm 16)$, with a 7\,$\%$ bias on the \cii\, APS due to the mask. At \zcii\,=7, power ratios are \cii \,/\,(residual CO)\,=\,$(2.7 \pm 1.9)$ and \cii \,/\,(residual \ci\,)=\,($9.4 \pm 6.6$), with an 8\,$\%$ bias on the \cii\, APS. For all redshifts, residual contamination becomes dominated by the faint sources for \rsigma\,$\geq 2$ since such sources are moderately affected by the mask.

\subsection{Field-to-field variance effects on residuals \label{sect:ftf_var}}

The detection of the \cii \, power spectrum depends on its contrast compared to that of the (masked) faint sources, the latter dominating the residual interloper contamination as shown in the previous section. In addition, these measurements will also be affected by the FtF variance. Indeed, G22 shows that FtF variance cannot be neglected when making a forecast for a LIM experiment, especially for the \cii \, LIM survey conducted by CONCERTO on a field the size of COSMOS.
We explore this here in more details.
We use the Uchuu simulation and the variance model from G22 (their Eq.\,8) to determine the faint CO and faint \ci\, APS variance. This equation is used to compute a factor $f$ to rescale the variance $\rm \sigma$ for a $\rm \Omega$\,=\,1\,$\rm deg^2$ field to the $\rm \Omega$\,=\,2\,$\rm deg^2$ field of the SIDES-Bolshoi simulation, assuming that the mean power spectrum $\rm \mu$ does not change between the two field sizes:
\begin{equation}
    f = \frac{\rm \sigma(\Omega=2deg^2)}{\sigma(\Omega=1deg^2)} \,.
\end{equation} The values of $f$ for the clustering term and the shot noise term are similar at the 1\% level. 

\begin{figure*}
    \centering
    \includegraphics[width=\textwidth]{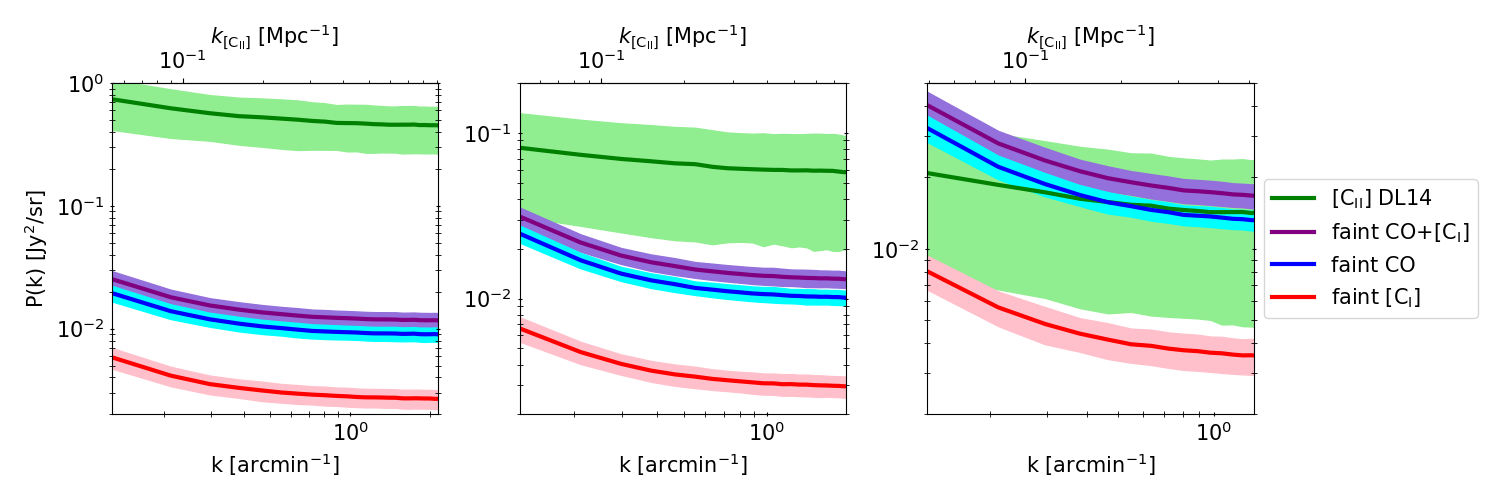}
    \caption{Mean APS (solid lines) and the 1\,$\rm \sigma$ FtF variance interval (coloured area) for a 2\,$\rm deg^2$ field, for \cii\,, faint CO, faint \ci\,, and the sum of the last two (green, blue, red, and purple, respectively). This is shown, from left to right, for 300-305\,GHz centred on \zcii \,=\,5.2, for 260-265\,GHz centred on \zcii \,=\,6.5, and for 235-240\,GHz centred on \zcii \,=\,7.}
    \label{fig:variance}
\end{figure*}

Figure\,\ref{fig:variance} shows the mean APS and its associated 1$\rm \sigma$ FtF variance interval in a 2\,$\rm deg^2$ field, for \cii\, and the faint interlopers at \zcii\,=5.2, 6.5, and 7. The \cii\, power amplitude varies by 40\% at \zcii\,=\,5.2 and by 60\% at \zcii\,=\,6.5 and 7. The faint (CO + \ci\,) APS varies by only 12-15\,\% from \zcii\,=\,5.2 to 7. Thus, the variance of faint sources does not have a strong impact on the contrast between residuals and \cii\, in comparison to the \cii\, variance. Table\,\ref{res_variance} lists the level of faint interloper contamination and its ratio to the $1\rm\sigma$ value of the \cii\, FtF variance. As a result, residual contamination is smaller than 20\,$\rm \%$ of the \cii \, FtF variance up to 287\,GHz (\zcii\,=\,5.62) and smaller than 50\,$\rm \%$ of the FtF variance up to 253\,GHz (\zcii\,=\,6.5). 

\begin{table}
\centering
\caption[]{Ratio (in percent) of faint interloper contamination (i.e. `unmasked faint CO+\ci\,') to the \cii \, power amplitude when the latter is evaluated at the mean and at the edges of the \cii\, $\rm \pm 1\sigma$ FtF variance interval. The last column gives (also in percent) the ratio between the faint interloper contamination and the $1\rm \sigma$ value of the \cii\, FtF variance. \label{res_variance}} 
\centering
$$
\begin{tabular}{c|ccc|c }
\hline
\hline
\noalign{\smallskip}
Frequency  & at +1\,$\sigma$ & at mean & at -1\,$\sigma$ & faint interlopers \\
      band &                 &         &                 & / 1\,$\sigma$ \\

\noalign{\smallskip}
\hline
\noalign{\smallskip}
 305\,GHz & 2\% & 3\% & 4\% & 7\%   \\
 287\,GHz & 6\% & 8\% & 12\% & 21\%   \\
 253\,GHz & 17\% & 27\% & 66\% & 45\%   \\
 237\,GHz & 87\% & 139\% & 348\% & 232\%   \\

\noalign{\smallskip}
\hline
\end{tabular}
$$

\end{table}



\section{Summary and conclusion \label{ccl}}

We tested a complete foreground deconfusion process for the \cii \, LIM survey conducted with CONCERTO using realistic mock data of the sub-millimetre sky provided by SIDES-Bolshoi (B17, B22). To build the SIDES mock data, a galaxy catalogue complete down to $10^7 \rm M_{\odot}$ and $z=7$ was first obtained using abundance matching between stellar mass and DM halo mass in a simulated DM light cone. Galaxy properties were then derived from their type (MS or SB), redshift, and stellar mass using empirically calibrated relations with appropriate scatters. Finally, mock cubes for CIB, CO, \ci\,, and \cii \, were produced. A foreground-contaminating mock cube was created by adding the separate foreground cubes to the \cii \, cube. \\

We first assessed the contamination by CIB (continuum) fluctuations. We tested the ability of PCA and arPLS to subtract the continuum emission for each \los, by exploiting its (electromagnetic) frequency-coherent distinctive feature.
Principal component analysis  is not able to separate the continuum from molecular and atomic lines up to \zcii\,=\,7. The arPLS method fits all the frequency-coherent emissions with a precision of 0.3\,$\rm \%$ in the mock spectra and sufficiently removes the continuum contamination for the two \cii\, models provided by SIDES up to $z=7$. For $5.2 \leq z \leq 7$, the residual CIB power amplitudes are lower than that of our fiducial DL14 (L18) \cii \, model by a factor of 72 (4). 

For the deconfusion with the line interlopers, we relied on a masking approach that uses an external COSMOS stellar mass catalogue, which provides accurate sky positions and redshifts for foreground sources to mask interloper-contaminated voxels. We show that $M_*$ is the deepest CO and \ci\, proxy available in the COSMOS field. We also investigated its use as a CO proxy. The most contaminating sources ($M_*$>$10^{8.1} \rm M_{\odot}$) are the least numerous and are above the L16 completeness relation, allowing us to efficiently remove the interloper contamination with a limited loss of the surveyed area. The spatial extent of the masks is defined with a (frequency-evolving) beam width criterion. To test the masking on our mock data in a realistic way, we masked only sources above the COSMOS (deep) completeness relation described in L16. 

For the SIDES-Bolshoi simulation, the power amplitude from bright CO and  \ci\, sources was lowered by four orders of magnitudes. Masking is also more efficient than expected in B22 as it also removes a fraction of the faint source signal at the same time, thanks to the clustering. The residual power amplitude left after masking is lower than $\rm 2 \times 10^{-2} Jy^2/sr$, rather independent of frequency, and is dominated by the faint sources that are not masked (especially CO sources). The CO power amplitude is lowered by a factor of 50 at $z=6.5$ with our masking, while it is lowered by a factor greater than 100 with a smaller masking depth in S18, indicating that the signal of sources below the mass threshold is stronger in SIDES than in S18. 

In the framework of our model, which contains a strong decrease in the \cii\, APS with redshift, we obtained the following results for the \cii\, APS measurement.

At \zcii\,=5.2, the power ratio is \cii\,/(residual CO+\ci\,)\,=\,$62 \pm 32$ (2.5\,$\rm \%$ residual contamination) for a 22\,\% surveyed area loss and 1\% of bias due to the mask. At this given redshift, the \cii\,  power spectrum varies by about 40\% due to the FtF variance. So, FtF variance is larger than any effect introduced by the masking or by residual contamination (which is $\rm \leq$ 4\% of FtF variance). 

At \zcii \,=\,6.5, the residual contamination after masking is a factor of \cii\,/(residual CO+\ci)\,=\,\,$5.5 \pm 3.6$ below the \cii\, power spectrum amplitude. This residual contamination and the reconstruction by \poker\, (see Appendix\,\ref{poker_appendix}) led to an overestimate of the \cii\, power spectrum by a factor of 1.4 for a 29\,$\rm \%$ loss of survey area. The variance estimate from SIDES-Uchuu gives a residual contamination by the faint sources that should not exceed half of the \cii\, FtF variance at this redshift. At \zcii \,=\,7, SIDES predicts a drop in \cii\, power spectrum amplitude by a factor of 16 with respect to \zcii \,=\,5.2 and an important FtF variance for a COSMOS-like field. Indeed, the measured residual contamination after masking is only a factor of $2.0 \pm 1.4$ below the \cii\, power spectrum amplitude.
 
Hence, residual contamination is still important above \zcii\,>\,6.5, despite the fact that most (>98\%) of the CO power has been masked. Masking even more sources to bring residues to a fainter level is tricky because the \cii\, APS is already slightly affected by the mask (i.e. underestimated by 7-8\% between $z$=\,6.5 and 7; see Appendix\,\ref{poker_appendix}) and the cost per masked source in a surveyed area is lower frequencies (i.e. higher redshifts). To reach the \cii\, signal at these redshifts, other or complementary methods for masking will be needed. \\

Our analysis makes use of the SIDES simulations \citep{bet22}, in which the \cii\, power spectrum is predicted with a much lower amplitude than in earlier models \citep[e.g.][]{Serra16}. For the first generation of LIM experiments, such as CONCERTO, and given such a low-amplitude \cii\, power spectrum, we expect the instrument noise to dominate. 
Van Cuyck et al. (in prep) will introduce the noise generator of SIDES and investigate the effects of instrument white noise on different LIM observables, such as auto-power spectra, cross-power spectra between lines and galaxies, and cross-power spectra between different lines within the LIM dataset. We have shown that our method for removing the continuum emission does not create any systematic residuals, but this will have to be tested with instrument noise. On the other hand, we expect masking to be robust to noise contamination because the interlopers' signal is removed regardless of the noise level. The robustness of cross-correlation measurements between different lines within the LIM dataset in the presence of instrument noise is also worth investigating in the context of component separation. 
Such a cross correlation could allow for a direct estimate of interlopers' residual power amplitude after masking. 
The last effect that would need further checks for the masking technique is the inclusion of the catalogues' redshift uncertainties, reported to be $\delta z/(1+z)$ = 0.025 below $z \leq 3$ for the COSMOS field \citep{cosmos2020}.

Finally, the CO-line emission, while being a strong foreground for the \cii\, survey, is of great importance. The CO part of the CONCERTO survey can be used to probe the CO SLED and the gas content of galaxies at z$\leq$3, including at cosmic noon (2\,$\leq$\,z\,$\leq$\,3), when star formation in galaxies statistically reaches its peak. The component separation methods and tools developed in this work for \cii\, can be adapted and used to exploit CO surveys. 

\begin{acknowledgements}
This project has received funding from the European Research Council (ERC) under the European Union’s Horizon 2020 research and innovation programme (grant agreement No 788212) and from the Excellence Initiative of Aix-Marseille University-A*Midex, a French “Investissements d’Avenir” programme. M.A acknowledges support from FONDECYT grant 1211951, CONICYT + PCI + Max Planck Institute for Astronomy MPG 190030, CONICYT + PCI + REDES 190194 and ANID BASAL project FB210003.

\end{acknowledgements}

%
%

\bibliographystyle{aea} 
\bibliography{masking} 

\appendix
\section{Measuring the APS of a masked distribution of point sources with POKER}
\label{poker_appendix}

In this paper we make extensive use of \poker, an algorithm developed to compute the APS of a masked 2D signal on a few square degrees at arbitrary high angular resolution \citep{poker}. \poker\, was developed to measure the power spectrum of the CIB in the context of experiments such as the infrared astronomical satellite (IRAS) and \textit{Planck} \citep{2011A&A...536A..18P} for which the CIB could be seen as an ubiquitous diffuse signal and treated as a random Gaussian field. From the direct Fourier transform of the masked data (pseudo-power spectrum), \poker\ inverts the mask induced aliasing and provides an estimate of the underlying APS that is unbiased on average when the signal is independent of the mask. When the signal of interest is highly non Gaussian such as in this work, with point sources and voids in between, the outcome of the power spectrum derived on masked area must however be reconsidered, and this, actually regardless of the APS estimator. 

\begin{figure}[htb]
    \centering
    \includegraphics[width=0.45\textwidth]{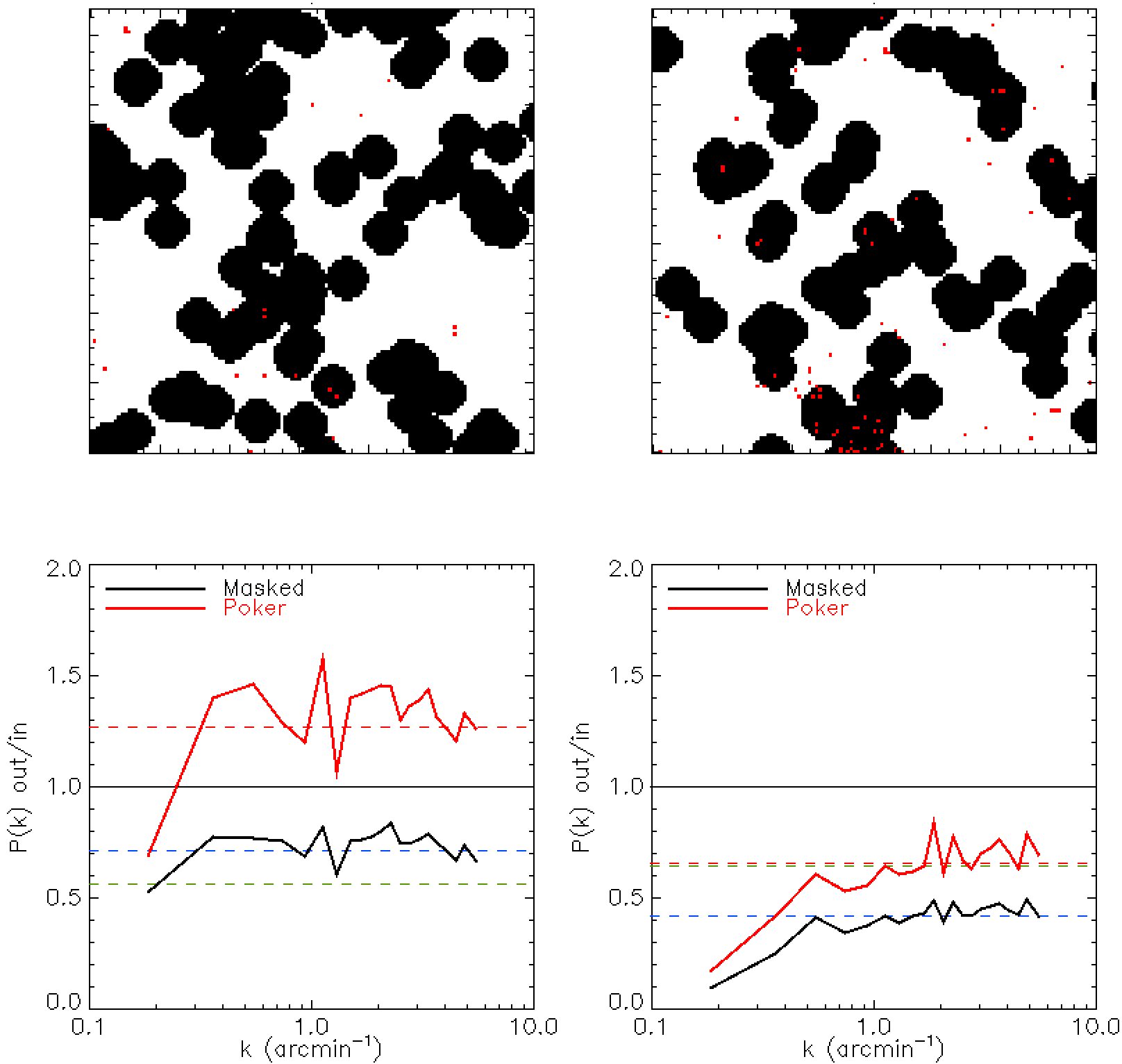}
    \caption{Illustration of the \poker\ correction for two different 114 arcmin$^2$ fields.  \emph{Top row}: CO+\ci\, masks corresponding to the two fields, as well as the \cii\, sources (red points). \emph{Bottom row}: APS of the masked \cii\, source maps, with or without correction of mask aliasing (labelled `Poker' and `Masked', respectively). 
    On the left-hand side, the mask covers 44\% of the sky patch (dashed blue line) but masks only 29\% of the \cii\, sources (dashed green line). \poker\ returns an output power spectrum that has the correct shape (flat, up to the sample variance) but that is overestimated by a factor of (1-0.29)/(1-0.44)=1.27. This is the dashed red line, and it matches the average level of the output APS. On the right-hand side, the mask covers only 36\% of the sky patch but 58\% of the sources. Hence, the output APS returned by \poker\ is underestimated by (1-0.58)/(1-0.36)=0.65.} 
    \label{fig:pk_mask_flux_ratio}
\end{figure}

For pedagogical purpose, we can start with the simplest example: a signal whose APS is constant for all angular modes. This APS can be that of a single point source or that of uniform white noise. In the case of uniform white noise, even a mask as complex as those shown in this work will leave some signal on the map from which the pseudo-power spectrum can be derived. In the case of a point source, if the source is masked, the pseudo-power spectrum is then zero and no measurement is possible. In this work, the signal can be approximated as resulting from a collection of $N_s$ point sources. If the sources had all the same flux, the intrinsic power spectrum would simply be $N_s$ times the uniform APS of single source of flux $\phi$. The mask covers a fraction $f_{mask}$ of the field but suppresses a fraction of sources $f_{sources}$ that is different because sources are not uniformly distributed, and loosely speaking, several sources can fall in the same hole. In this case, the pseudo-power spectrum is underestimated by $f_{sources}$ and corrected up by $f_{mask}$ by \poker. Figure~\ref{fig:pk_mask_flux_ratio} illustrates two different examples where the mask and the source locations are taken directly from a $114 \,\rm arcmin^2$ section of the SIDES-Bolschoi simulation and their flux is fixed to 1: depending on the effective (a priori unknown) clustering of the sources compared to the mask, the reconstructed APS is larger or smaller than the intrinsic \cii\, APS. In this case, it is therefore impossible by construction to measure an unbiased APS in the presence of a mask, and the bias will depend on the clustering properties of the signal compared to the mask. We can extend this reasoning to the sum of several distributions of point sources, each distribution with a different flux. Each one of them will lead to an APS estimate biased by a different $f_{sources}/f_{mask}$ so that the final APS is some average of these ratios, weighted by the source flux. Last, the convolution by the point spread function adds another feature. Indeed, it gives a spatial extension to the initial point sources, so that their signal can be only partially masked and more degeneracies show up when computing the absolute level of the pseudo-power spectrum. It is therefore essential to test the APS derivations on realistic source and mask simulations to assess the magnitude of these potential biases. 

\begin{figure}[htb]
    \centering
    \includegraphics[width=0.5\textwidth]{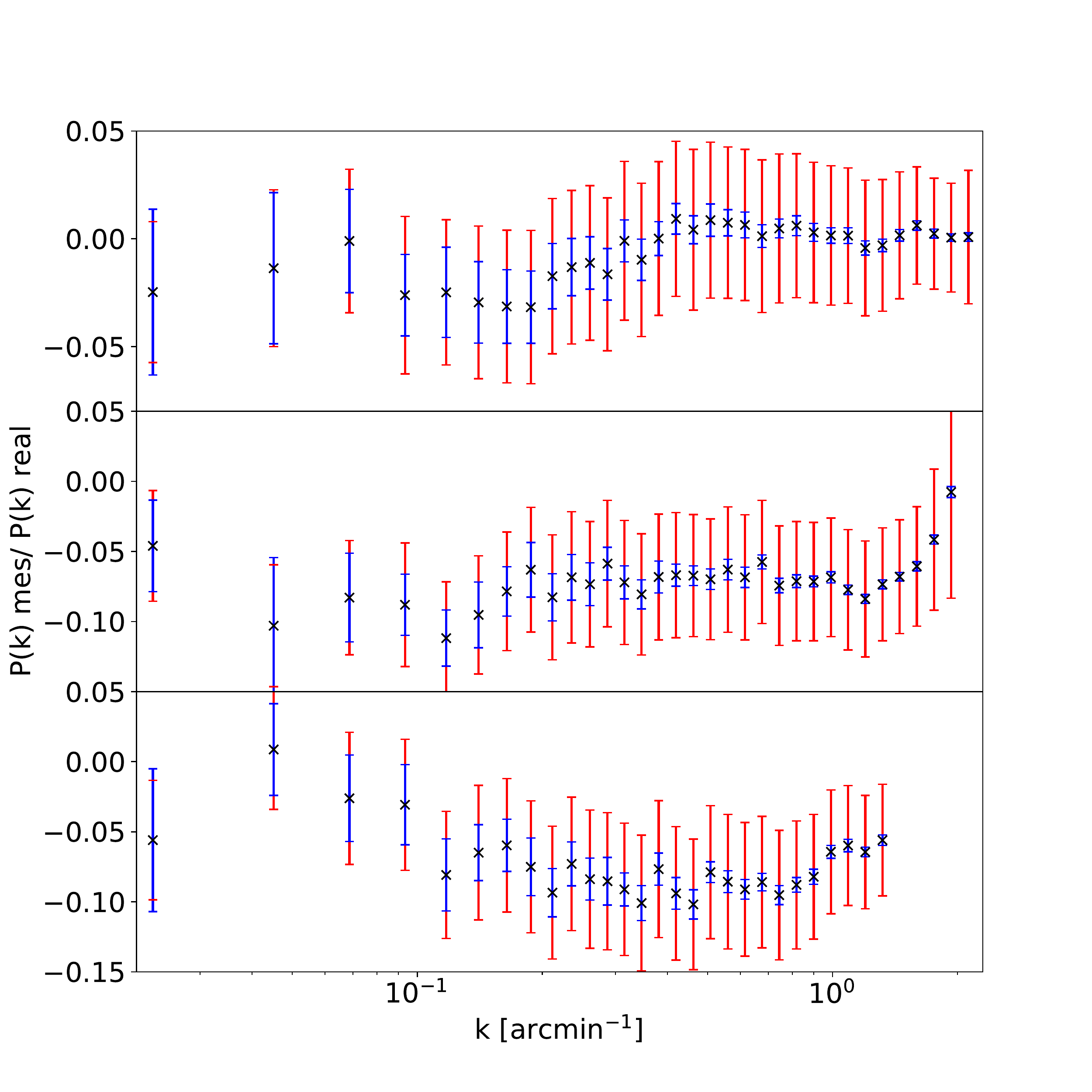}
    \caption{Average relative error (black crosses) between the \cii\, APS measured in the 54 Uchuu sub-fields masked with \rsigma\,=\,3 as a function of spatial frequency, $k$, at $z=5.2$ (top), $z=6.5$ (middle), and $z=7$ (bottom). The mean errors over the $k$ modes are <$1\rm \%$, $7\rm \%,$ and $8\rm \%$ for each redshift, respectively. The dispersion (red error bars) normalized by the square root of the number of realizations is also represented. The mean sizes of these normalized error bars over the $k$ modes are $3\rm \%$, $4\rm \%,$ and $5\rm \%$ for each redshift, respectively. For reference, the equivalent normalized error bars for a perfect Gaussian diffuse signal from the same number of Monte Carlo realizations are over-plotted in blue. } 
    \label{fig:uchuu_bias}
\end{figure}

Figure~\ref{fig:uchuu_bias} presents the results of this test, from which we built the error bars given in Sect.\ref{sect:result_masking}. For each of the 54 fields of $\rm 2\,deg^2$ cut in the Uchuu simulation, we built a mask with \rsigma\,=\,3 following Sect.\,\ref{sect:masking} and computed the ratio between the output (masked) and input \cii\, APS. We then computed the average of this ratio over all the 54 fields and show that it is not biased low by more than $1\rm \%$, $7\rm \%,$ and $8\rm \%$ on average for the three redshifts (\zcii \,=\,5.2, 6.5, and 7), up to an angular scale of $k_{lim}$ = 2.14, 1.94, 1.33\,\karcmin\,, respectively. The bias increases with redshift because the masked area is larger, as it depends on the beam size. Error bars for an APS obtained using a mask at a given frequency are the dispersion of this ratio at this frequency. The obtained errors bars represent 25\%, 33\%, and 34\% of the masked APS at \zcii \,=\,5.2, 6.5, and 7, respectively, and are also used for smaller \rsigma\, masks. Error bars displayed in Fig.\,\ref{fig:uchuu_bias} are normalized by the number of realizations. They are constant for all angular scales because they result from an effective overall scaling by the mask to flux ratio (see e.g. Fig.~\ref{fig:pk_mask_flux_ratio}) that varies from one field to another. For reference, the error bars that would come from sampling variance of a diffuse signal are displayed in blue and their amplitude decreases (as expected) with the increasing number of available angular modes at higher $k$.

\begin{figure}
    \centering
    \includegraphics[width=0.45\textwidth]{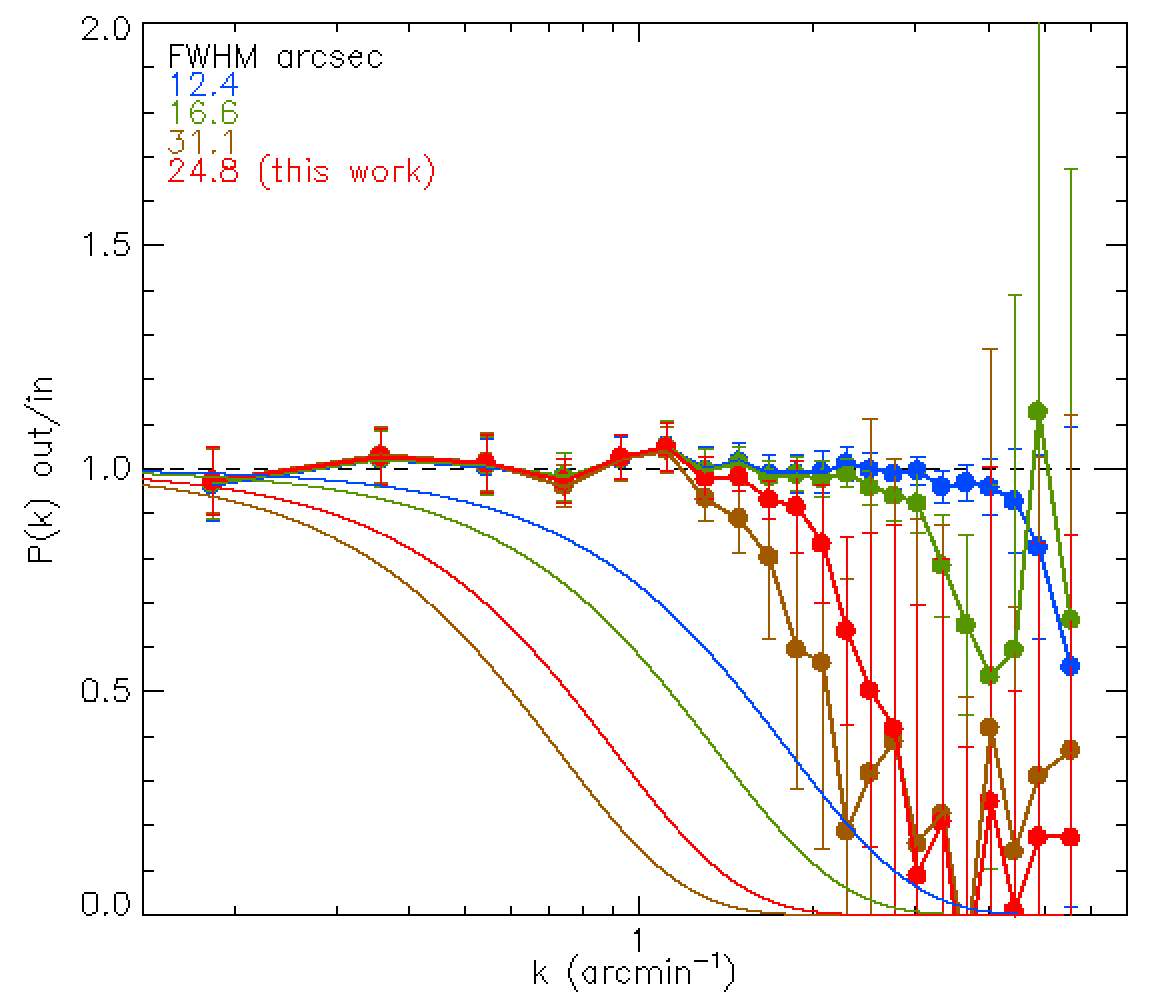}
    \caption{Output (masked) over input (intrinsic) \cii\, APS ratio for various beam sizes but with the same mask. The smaller the beam, the larger the  $k$ angular mode up to which the output APS equals the input one. The analytical beam is displayed in solid lines for reference and shows that the effect only appears at high angular scales, where the beam has mostly smeared out any signal.}
    \label{fig:subbeam_bias}
\end{figure}

At smaller angular scales, there is a systematic small-scale bias. In Fig.\,\ref{fig:subbeam_bias} we check via simulations that this bias either damps or boosts the estimate depending on the relative size of the mask and the FWHM. Table\,\ref{tab:k_lim} gives the spatial frequency at which this bias between the measured and the expected APS becomes larger than $\rm 20 \%$ and $\rm 50 \%$, in the interloper contaminated maps masked at \rsigma\,=\,3.0\,. These values are well above the beam cutoff, and for consistency, in all this work, we restrict to angular scales $k<k_{\rm 20\%}$.

\begin{table}
    \caption[]{ Spatial frequency (in \karcmin\ ) at which the pseudo power spectrum becomes biased by 20\,$\rm \%$ and 50\,$\rm \%$, and the corresponding angular size, $r$, as a fraction of the FWHM. 
    \label{tab:k_lim} }
\centering
\tiny{
\begin{tabular}{l|cc|cc}
\hline
\hline
\noalign{\smallskip}
frequency & $k_{\rm 20\%}$ & $r_{\rm 20\%}$ [FWHM] & $k_{\rm 50\%}$ & $r_{\rm 50\%}$ [FWHM] \\
\noalign{\smallskip}

\hline
\noalign{\smallskip}
305 GHz & 2.14 & 1.36 & 2.35 & 1.24 \\
253 GHz & 1.94 & 1.24 & 2.14 & 1.36 \\
237 GHz & 1.33 & 1.71 & 1.46 & 1.55 \\
\noalign{\smallskip}
\hline
\end{tabular}}

\end{table}

\end{document}